\PassOptionsToPackage{hyphens}{url}
\documentclass[mnsc,blindrev]{format/informs3}

\OneAndAHalfSpacedXI

\usepackage[authordate, natbib=true, backend=biber, dashed=false]{biblatex-chicago}
 \renewcommand*{\bibfont}{\small}

\usepackage{amsmath}
\usepackage{array}
\usepackage{graphicx}
\usepackage{caption}
\usepackage{subcaption}
\usepackage{siunitx}
\usepackage{multirow}
\usepackage[flushleft]{threeparttable}
\usepackage{pifont}       
\usepackage{hyperref}
\usepackage{url}

\RUNAUTHOR{}
\RUNAUTHOR{Zhang, Cui, and Zhang}
\RUNTITLE{The Impact of AI Search on the Online Content Ecosystem}
\TITLE{The Impact of AI Search on the Online Content Ecosystem: Evidence from Google and Reddit}
\ARTICLEAUTHORS{%
\AFF{}
\AUTHOR{Peibo Zhang}
\AFF{Goizueta Business School, Emory University, \EMAIL{peibo.zhang@emory.edu}}
\AUTHOR{Ruomeng Cui}
\AFF{Goizueta Business School, Emory University, \EMAIL{ruomeng.cui@emory.edu}}
\AUTHOR{Dennis J. Zhang}
\AFF{Olin Business School, Washington University in St. Louis, \EMAIL{denniszhang@wustl.edu}}
}

\ABSTRACT{
Search engines traditionally complement online content platforms by directing users seeking information to external websites. The emergence of generative AI search tools that summarize answers directly on the results page may disrupt this relationship by making visits to source platforms optional. We study this question using Google AI Overviews and Reddit, one of the largest online discussion platforms. Our identification exploits Google's content moderation policy: Safe-for-Work (SFW) Reddit communities are indexed by Google organic search and surfaced in Google AI Overviews, while Not-Safe-for-Work (NSFW) communities, though indexed by organic search, are prohibited from being referenced in AI Overview summaries. Using a difference-in-differences design, we find that AI Overviews increase engagement in SFW communities: daily comments rise by 12.0\% and the number of commenting users by 12.4\% relative to NSFW communities. The effects are concentrated in experience-based discussions (opinions, advice, and personal experiences) rather than fact-based information. However, the subsequent introduction of Google AI Mode, which allows users to interact conversationally with the AI summary, largely eliminates these gains in experience-based content. These results suggest that the effects of AI search depend critically on interface design and content type.
}

\begin{document}

\maketitle


\section{Introduction}\label{sec:intro}

The internet's content ecosystem depends on a long-term complementary relationship with search engines. Search engines index and surface content, directing users to the platforms that produce it. Content platforms, in turn, rely on this referral traffic to sustain engagement and grow. The recent rapid integration of generative AI into search is now reshaping this relationship at scale. AI search features such as Google AI Overviews synthesize answers directly on the search results page, satisfying users' information needs by providing a static summary without necessarily requiring a visit to the source website. As AI search continues to develop, it challenges the fundamental relationship between search engines and the online content ecosystem.

The transition from traditional search to AI search for information retrieval has involved three distinct stages, each with different implications for the content ecosystem. In the first stage, users relied on search engines, which indexed and ranked content, and clicked through to the source websites to find answers. In the second stage, standalone AI chatbots such as ChatGPT allowed users to obtain answers from static, historical training data. Because these tools operated outside the search ecosystem, users who obtained answers had no direct path back to a source website, which diverted traffic away from content platforms. In the third and current stage, search engines have begun embedding generative AI directly into the search results page. Unlike chatbots, AI search features such as Google AI Overviews are grounded in real-time web content and provide inline citations that link back to the original sources. This architecture preserves a direct pathway from the AI summary to the content platform, meaning AI search can both satisfy information needs and funnel users to the underlying content. Whether AI search substitutes for, sustains, or increases platform engagement is the central empirical question we address.

The emergence of AI search has raised concern and drawn intense attention from industry and legislators. By late 2025, dozens of lawsuits had been filed against AI providers over the use of publishers' content \citep{pressgazette2025lawsuits}, including a growing number that target AI search summaries specifically: Chegg v. Google \citep{chegg2025}, Penske Media v. Google \citep{penske2025}, and The New York Times v. Perplexity AI \citep{nyt2025perplexity}. Regulators have followed suit: U.S. senators urged antitrust investigations into generative AI features that summarize publisher content \citep{klobuchar2024}, and the European Parliament called for remuneration obligations on providers of general-purpose AI models that use publishers' copyrighted content \citep{europeanparliament2025}. Despite this intense scrutiny from industry and regulators, rigorous causal evidence on how AI search affects the content ecosystem remains scarce. 

To explore these significant yet understudied questions, we answer three core questions in this paper: (1) How does AI search affect user engagement in the online content ecosystem? (2) What mechanisms drive heterogeneous effects across different types of content? (3) How does the design of the AI search interface moderate its impact on the online content ecosystem?

We address these questions by studying the impact of Google AI Overviews on Reddit. We focus on Google because it processes over 90\% of global search traffic.\footnote{https://www.statista.com/statistics/1381664/worldwide-all-devices-market-share-of-search-engines/} Reddit is one of the largest online discussion platforms, with over 121.4 million daily active users \citep{reddit2026_10k}. Reddit is organized into topic-specific communities called ``subreddits,'' where users post submissions and comment on existing posts. Reddit provides a particularly informative setting for two reasons. First, Reddit's community structure provides a clean identification strategy: Safe-for-Work (SFW) subreddits are indexed by Google and surfaced in AI Overviews, while Not-Safe-for-Work (NSFW) subreddits are excluded under Google's content policies, which bar sexually explicit content from AI search features \citep{google2025aio,google2025content}. Second, Reddit's content ranges from factual reference communities to those centered on personal experience and interactive discussion, enabling heterogeneity tests within a single platform.

Using public data on Reddit user engagement from the Arctic Shift archive, we construct a balanced panel of \num{105012} public subreddits, each with at least 10 submissions in December 2023, and track their daily user engagement from January 2024 through July 2025 (\num{60696936} subreddit-day observations). We focus on two outcomes that capture demand-side user engagement: the number of daily comments, which counts all comments posted within a subreddit on a given day, and the number of daily unique comment authors, which counts all distinct users who commented at least once.\footnote{We also examine supply-side metrics (number of submissions and unique submission authors) but do not find statistically revealing results.} We set the treatment date to August 15, 2024, when AI Overviews expanded internationally to six additional countries \citep{budaraju2024}, thereby covering the majority of Reddit's global user base. We employ a difference-in-differences design with subreddit and day fixed effects.

Contrary to the substitution effect identified in prior literature, we find that AI Overviews cause a statistically significant \textit{increase} in Reddit engagement. Daily comments in SFW subreddits increase by 12.0\% and daily comment authors by 12.4\%, both relative to NSFW subreddits. The positive effects are broad-based across the outcome distribution and all community size quartiles, with small and mid-sized communities experiencing the largest relative gains, suggesting that AI search surfaces the ``hidden gems'' of the platform and directs attention to previously underexplored communities.

To explain why prior evidence suggests substitution while our results show increased engagement, we investigate how AI search affects different types of content. Online content broadly falls into two categories: fact-based content, such as encyclopedia entries and technical Q\&A, and experience-based content, such as personal opinions, subjective advice, and lived experiences. Drawing on the information goods literature \citep{nelson1970}, we classify fact-based content as search goods, where a user can evaluate the answer from a summary alone (e.g., ``What is the capital of France?''), and we classify experience-based content as experience goods, where the value comes from reading the full discussion (e.g., ``What is it like to live in Paris?''). We develop a framework in which AI search exerts two countervailing forces on these two content types. For search goods, AI summaries satisfy users' information needs directly, reducing visits to the source; we call this a \textit{replacement effect}. For experience goods, AI summaries surface content but cannot replicate its richness, driving users to the platform to consume it firsthand; we call this a \textit{discovery effect}.

We test this by classifying all subreddits as fact-based or experience-based using Google's Gemini 3 Flash large language model under the original \citet{nelson1970} taxonomy. Our results confirm that experience-based communities exhibit treatment effects 2.3 times larger for comments and 2.8 times larger for comment authors than those of fact-based communities. These findings support our framework: AI search benefits experience-oriented platforms substantially more than fact-oriented ones.

If the discovery effect operates because a static AI summary cannot fully satisfy users' needs for experience-based content, then a more interactive AI search interface that better addresses those needs might weaken it. We further test this by comparing Google AI Overviews (static summaries) with Google AI Mode (conversational interface), which allows users to ask follow-up questions within the search page itself. We find that the experience-based premium is sharply attenuated after the launch of AI Mode: it turns slightly negative for comments and falls by 59\% for comment authors while remaining positive, indicating that a conversational interface largely absorbs the depth of redirected engagement and community participation. The interactive AI search interface thus satisfies more information needs even for experience-based content, suggesting that the magnitude of the discovery effect also depends critically on AI search interface design. We further show that this attenuation is not driven primarily by alternative explanations, such as the maturation of AI search technology or growing user trust in AI-generated results over time.


Our findings make several important contributions. Academically, we provide some of the first empirical evidence on how AI search affects the online content ecosystem. In contrast to the substitution effect documented for AI chatbots, we show that AI search can increase engagement. We develop a framework distinguishing a replacement effect for fact-based content from a discovery effect for experience-based content, and confirm that the discovery effect dominates on discussion-oriented platforms. We further show that this effect is not fixed but depends on interface design, attenuating sharply when the search interface shifts from a static summary to a conversational format. For practitioners and policymakers, our results show that AI search does not uniformly harm the platforms it summarizes and can even raise engagement for discussion-oriented content, evidence that speaks directly to the ongoing lawsuits between content platforms and AI providers. They also caution legislators against one-size-fits-all rules, since the effect of AI search is not uniform but hinges on the type of content a platform holds and the interface the search engine adopts. 



\section{Literature Review}\label{sec:literature}

Our study connects three streams of research: one on how artificial intelligence reshapes the online content ecosystem, one on substitution and complementarity in information intermediation, and one on how AI reshapes the operations of digital platforms.

\subsection{AI and the Online Content Ecosystem}\label{sec:lit_ecosystem}

A growing body of work asks whether generative AI substitutes for or complements the platforms whose content it draws on. Early evidence, focused on AI chatbots, mostly points toward substitution. \citet{burtch2024} find that ChatGPT reduced Stack Overflow activity, while Reddit showed no comparable decline, which they attribute to Reddit's stronger social fabric. \citet{delriochanona2024} document a 25\% drop in Stack Overflow questions within six months of ChatGPT's release. \citet{padilla2025} show that LLM adoption substitutes for traditional search, cutting total searches by roughly 20\%, with browsing losses concentrated on smaller websites rather than large ones, while \citet{gholami2026} find that LLM adoption increases the number of distinct websites users visit, suggesting complementarity at the browsing level. 

AI chatbots differ from AI search in two ways that matter for the content ecosystem. First, a chatbot is a standalone destination that users visit instead of a search engine, whereas AI search is embedded in the existing search workflow, surfacing automatically above the organic results for queries users already issue rather than requiring separate adoption. Second, a chatbot generates a self-contained answer and, in its basic form, does not point back to the sources it draws on, whereas AI search cites the sources it summarizes, preserving a click-through path to the originating platform. Because a chatbot neither rides the existing search workflow nor links back to its sources, it tends to substitute for visits to the underlying platforms. The lessons learned from AI chatbots may therefore not apply directly to AI search. To the best of our knowledge, only two recent papers provide empirical evidence on its effect on content platforms: \citet{khosravi2026} estimate that Google AI Overviews reduce Wikipedia pageviews by about 15\%, and \citet{agarwal2026} find in a field experiment that AI Overviews cut outbound organic clicks by 38\%.

We also study successive generations of AI search, moving from the static summaries of AI Overviews to the conversational interface of AI Mode, which connects our work to a broader literature on technology advancement and user trust over time. \citet{nortonbass1987} model how successive generations of a high-technology product reshape adoption, with the newer generation substituting for the older. A parallel literature shows that trust in an automated system calibrates to its perceived capability and usefulness \citep{leesee2004, davis1989} and updates as users observe the system perform \citep{dietvorst2018}. This raises an alternative to the interface channel: a rise in trust as AI capability matures, rather than the interface itself, could change how users engage with the source. By comparing two generations of AI search, we separate the effect of the interface from this contemporaneous rise in trust.


We contribute to this stream of literature in three ways. First, the platforms studied in prior work, Stack Overflow and Wikipedia, host predominantly fact-based content that AI can readily summarize; we study a platform spanning both fact-based and experience-based content, enabling heterogeneity tests within one identification framework. Second, we find that AI search increases rather than decreases engagement, showing that the discovery channel can dominate even for comprehensive AI summaries. Third, we explore multiple factors that shape the effect of AI search, specifically the type of content a platform hosts and the design of the AI interface that mediates discovery.

\subsection{Substitution and Complementarity in Information Intermediation}\label{sec:lit_search}

We also connect to a literature on how third-party aggregators reshape traffic to content producers. \citet{jeon2016} formalize the dual nature of an aggregator: it can steal business from publishers by providing a self-contained substitute, and it can also expand readership by lowering the cost of discovery. \citet{atheymobius2021} use the December 2014 shutdown of Google News in Spain to estimate a sizable readership-expansion effect, finding that overall news consumption fell by roughly 20\% when the aggregator was removed. Studying the October 2014 opt-in policy in Germany, \citet{calzada2020} find more muted effects, with a significant decline concentrated among outlets that opted out. \citet{chiou2017} exploit the temporary removal of Associated Press content from Google News and find that the snippet-based aggregator on net increased visits to publisher sites, reinforcing the readership-expansion effect. 

We contribute to this literature by extending the substitution-complementarity question from traditional news aggregation to AI-generated search summaries. AI Overviews represent a qualitative shift from traditional snippets: they synthesize complete answers rather than displaying a sentence or two. Our findings show that even with this more comprehensive form of aggregation, the complementarity result persists for experience-based content, where AI summaries cannot replicate the richness of the original discussion.

\subsection{AI and Platform Economics}\label{sec:lit_platform}

Our study connects to a broader literature on how generative AI reshapes the operations of firms and platforms, including how it reshapes supply chains \citep{cohen2026, song2026ai}, raises the productivity of knowledge workers as a general-purpose technology with wide task exposure \citep{noy2023, peng2023, dellacqua2023, eloundou2024}, makes operational decisions rather than merely assisting with them \citep{cohen2025wages, jiang2025ai}, and shapes how users respond to and rely on AI through the design of its interface rather than its mere presence \citep{xu2024, dossantosdisorbo2025}. Prior literature also shows that operational transparency, making the work and sources behind a result visible, raises its perceived value and users' willingness to engage \citep{buell2017, buell2011}. This is the mechanism by which AI search, which cites and links to the sources it summarizes, can channel users to the underlying platform.

Because AI search stands between users and the platforms it summarizes, its effect also turns on the economics of platforms and intermediaries, where value depends on how an intermediary connects distinct groups and on whether it competes with or complements the parties it links \citep{parker2005, cohen2022, zhuo2017, ding2022quality}. The central question is whether such mediation concentrates or broadens activity \citep{fleder2009, brynjolfsson2003, aparicio2025}, particularly for user-generated content, whose value reflects the community that produces it \citep{ransbotham2011, zhang2011, wang2023mindgap} and resists replication by an AI summary.


We connect these strands by studying how AI mediation reshapes engagement on a major content platform, showing that the sign and size of the effect depend jointly on the type of content the platform hosts and on the design of the AI interface that mediates discovery. Consistent with AI search acting as a discovery channel that broadens rather than concentrates activity, small and mid-sized communities gain the most. This finer-grained, operations-oriented view is essential for predicting which platforms benefit and which are displaced as AI mediation expands.

\section{Empirical Setting and Data}\label{sec:data}

In this section, we introduce the empirical setting and the datasets.

\subsection{Empirical Setting}\label{sec:setting}

\begin{figure}[t!]
\centering
\caption{Illustration of Google AI Overviews}
\label{fig:aio_screenshot}
\vspace{0.1in}
\hfill
\begin{subfigure}[t]{0.33\textwidth}
\centering
\includegraphics[width=\textwidth]{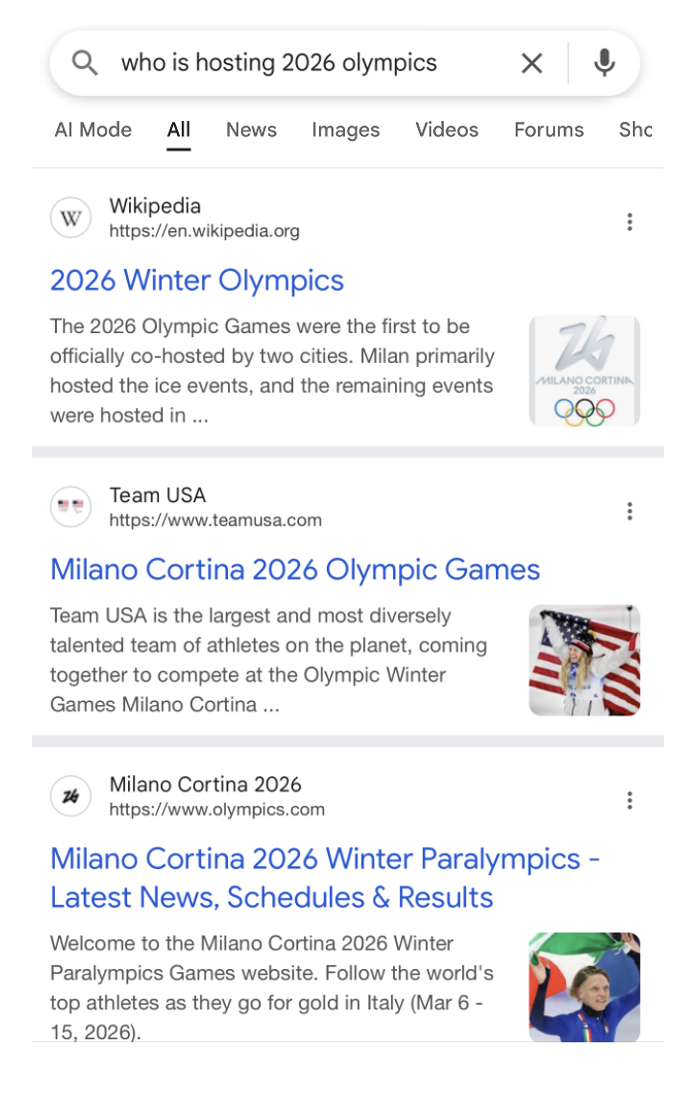}
\caption{Without AI Overview}
\label{fig:aio_without}
\end{subfigure}
\hfill
\begin{subfigure}[t]{0.5\textwidth}
\centering
\includegraphics[width=\textwidth]{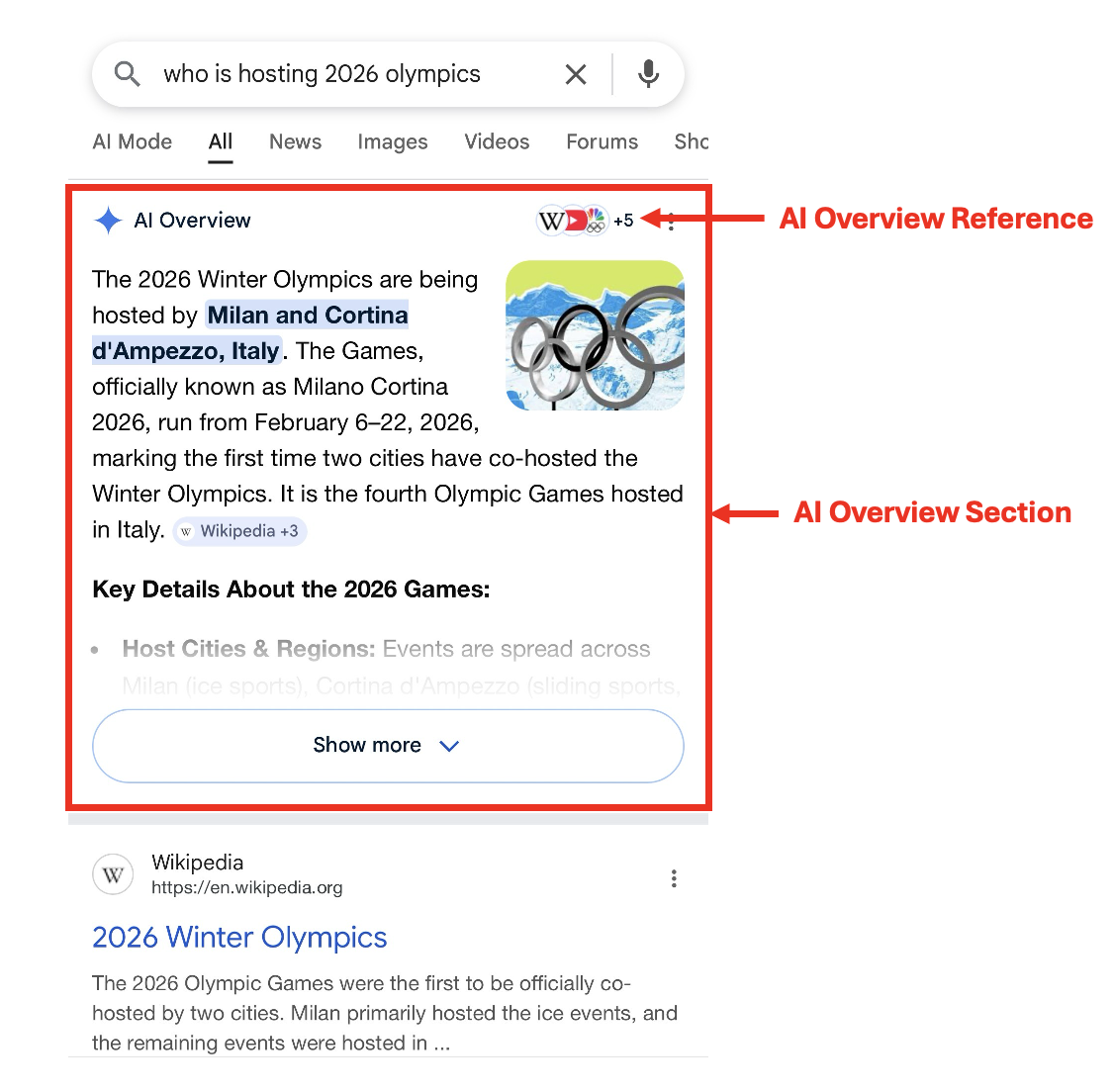}
\caption{With AI Overview}
\label{fig:aio_with}
\end{subfigure}
\hfill\mbox{}
\vspace{0.05in}
\parbox{0.95\textwidth}{\scriptsize \textit{Note}: Both panels show Google Search results for the query ``who is hosting 2026 olympics.'' Panel (a) displays a search results page without an AI Overview, showing only traditional organic search results. Panel (b) displays a search results page with the AI Overview feature enabled: the AI Overview section (red box) presents a generative AI summary synthesized from multiple sources, with inline citations to referenced websites.}
\end{figure}

We study the impact of Google AI Overviews on Reddit. Google AI Overviews is a feature integrated into Google Search that uses generative AI to synthesize answers directly on the search results page. As illustrated in Figure~\ref{fig:aio_screenshot}, rather than returning a ranked list of links, AI Overviews generate a concise summary that addresses the user's query, with inline citations linking to the source websites. Google rolled out AI Overviews to all U.S.\ users in May 2024 and completed the first international expansion on August 15, 2024, bringing the feature to the United Kingdom, India, Japan, Indonesia, Mexico, and Brazil \citep{budaraju2024}.


\begin{figure}[t!]
\centering
\caption{Illustration of the Reddit Interface}
\label{fig:reddit_screenshot}
\vspace{0.1in}
\includegraphics[width=0.85\textwidth]{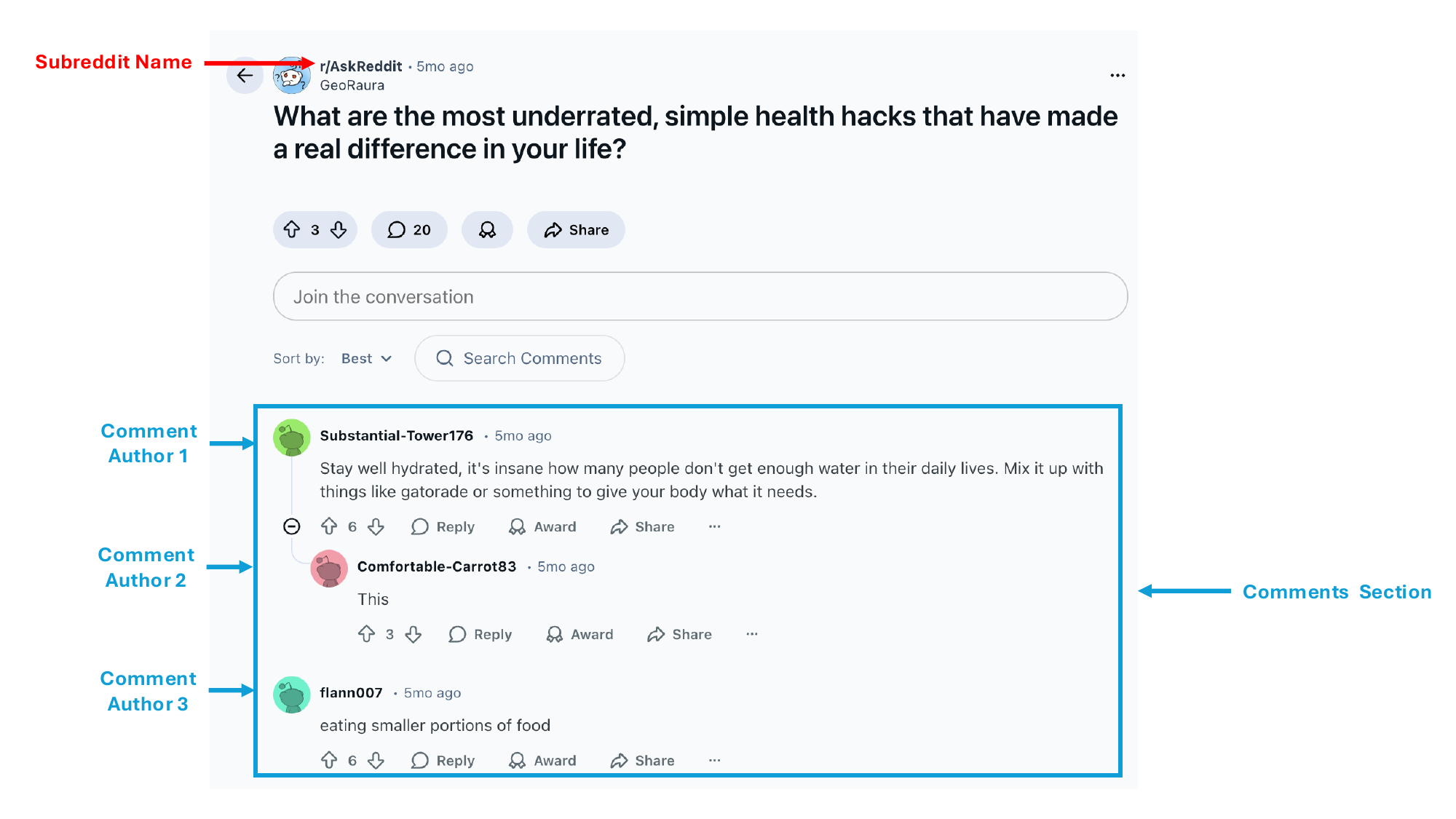}
\vspace{0.05in}
\parbox{0.85\textwidth}{\scriptsize \textit{Note}: This figure shows a submission and its discussion thread in the subreddit r/AskReddit. The \emph{submission} is a question posted by a submission author. The \emph{comment section} contains responses from three distinct comment authors.}
\end{figure}

Reddit is one of the largest online discussion platforms, with over 121.4 million daily active users as of 2025 \citep{reddit2026_10k}. The platform is organized around subreddits, which are community-specific forums that each focus on a particular topic, interest, or theme (e.g., r/science, r/cooking, r/personalfinance). Communities range from broad general-interest forums with millions of subscribers (e.g., r/AskReddit) to highly specialized niche communities. Figure~\ref{fig:reddit_screenshot} illustrates the interface of a typical subreddit. Users can create submissions (also called ``posts'') within a subreddit, and other users can write comments on these submissions. Comments are nested hierarchically: users can reply to the original submission or to other comments, producing branching conversation threads. Comments and the number of unique commenting authors represent the primary forms of demand-side user engagement on the platform.

Reddit is a particularly suitable setting for studying the impact of AI search features. First, Reddit's content spans a wide spectrum, from experience-sharing communities centered on social interaction, opinion exchange, and community engagement (e.g., r/relationship\_advice) to subreddits that serve as quick-answer repositories for factual questions (e.g., r/learnpython). Second, Reddit's subreddit structure provides a natural partition between communities exposed to Google AI Overviews and those that are not. Reddit communities include both ``Safe-for-Work'' (SFW) subreddits, which contain general-audience content, and ``Not-Safe-for-Work'' (NSFW) subreddits, which contain adult or sensitive material. Under its content moderation policy, Google indexes NSFW content in organic search results but prohibits AI Overviews from referencing it. SFW content, by contrast, appears in both organic search results and AI Overview summaries \citep{google2025aio,google2025content}. This creates a natural treatment group (SFW communities) and control group (NSFW communities). We validate the enforcement of this policy through a manual audit of Google searches in Appendix~\ref{sec:appendix_aio_validation} and confirm that AI Overviews reference only SFW content.


Both groups share the same Reddit platform infrastructure, user interface, and recommendation algorithms, and are subject to common platform-wide shocks, satisfying the key requirement for a credible control group in a difference-in-differences design.\footnote{\citet{corradini2021} document that approximately 98\% of Reddit authors who participate in NSFW communities do not appear in SFW communities, confirming that these two populations are effectively disjoint.}


\subsection{Data}\label{sec:panel}

We obtain Reddit data from the Arctic Shift archives \citep{heitmann2024arcticshift}, a comprehensive collection of publicly available Reddit user engagement data that succeeds the earlier Pushshift project. We define our analysis cohort using subreddit activity in December 2023, a baseline month that predates the broad rollout of Google's AI search features.\footnote{Google began testing AI-generated search results (then called Search Generative Experience) as an opt-in Search Labs experiment in May 2023 (\url{https://blog.google/products/search/generative-ai-search/}) and extended the test to a small share of U.S.\ users who had not opted in in March 2024 (\url{https://www.engadget.com/google-will-start-showing-ai-powered-search-results-to-users-who-didnt-opt-in-093036257.html}), but the feature was not rolled out broadly until May--August 2024.} We retain all subreddits with at least 10 submissions during December 2023. From this set we exclude subreddits that fall outside the treatment mechanism: user-profile pages (those with the \texttt{u\_} prefix), which are single-author personal pages rather than community discussions; quarantined subreddits, which Google does not index and therefore cannot surface in AI Overviews; and non-public subreddits (private, restricted, or otherwise gated), which are not visible to Google's crawler. The resulting cohort contains \num{105012} subreddits. All of our main results remain robust to alternative submission thresholds (Appendix~\ref{sec:appendix_cohort}).

We classify each subreddit's NSFW status based on its content during the baseline month of December 2023: a subreddit is labeled NSFW if more than 50\% of its submissions are flagged as ``over 18'' by either the subreddit's moderators or Reddit's content classification system. This classification is fixed at the baseline month and does not change over the analysis period. Our primary cohort of \num{105012} subreddits comprises \num{70683} SFW subreddits and \num{34329} NSFW subreddits.

For each subreddit, we track user engagement from January 1, 2024 to July 31, 2025. For each subreddit-day, we compute two daily metrics: (1) the number of comments, defined as all comments published under any submission or in reply to other comments on a given day; and (2) the number of unique comment authors, defined as all distinct users who posted at least one comment on a given day. We assemble a fully balanced panel of all cohort subreddits and all calendar days in our study period. Days on which a subreddit had no observed activity receive a count of zero. The resulting panel contains \num{60696936} subreddit-day observations.\footnote{\num{105012} subreddits $\times$ 578 days = \num{60696936}.}

\begin{figure}[t!]
\centering
\caption{Model-Free Evidence: Average Daily Activity by Treatment Group}
\label{fig:model_free}
\vspace{0.1in}
\includegraphics[width=\textwidth]{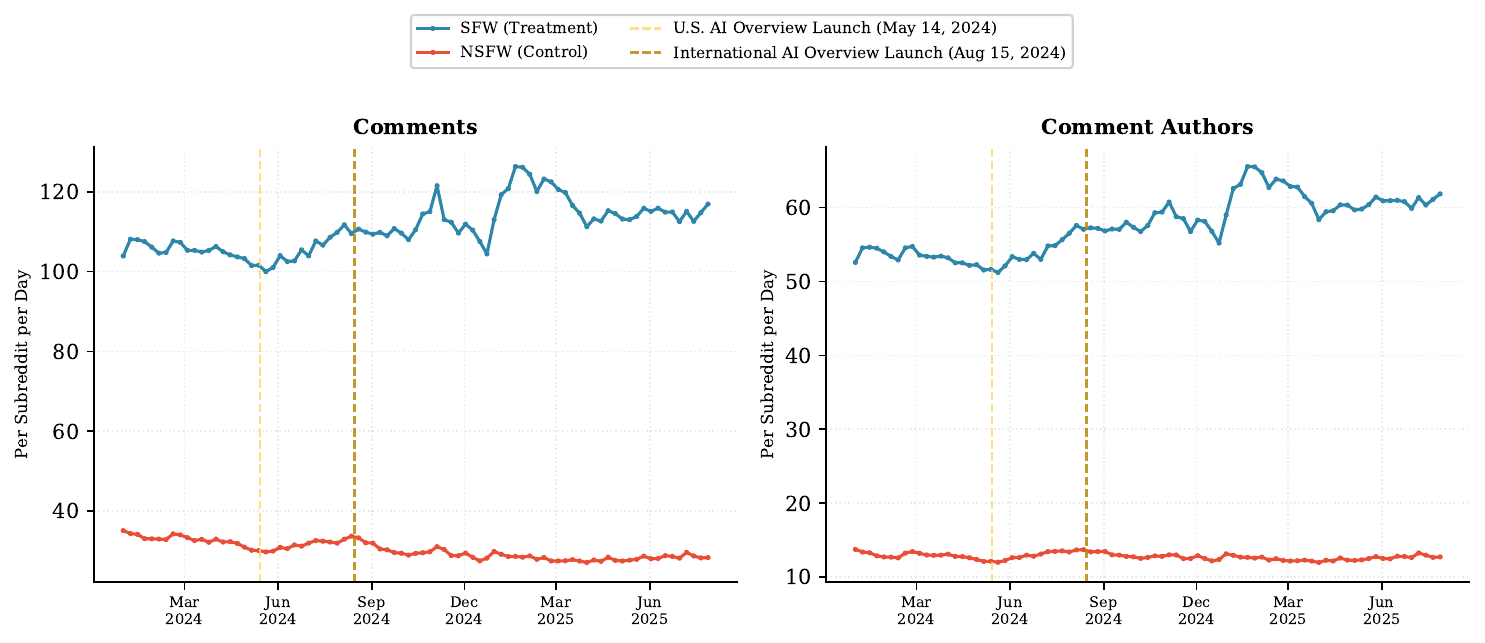}
\vspace{0.05in}
\parbox{\textwidth}{\scriptsize \textit{Note}: Each panel plots the average daily outcome across all subreddits in the group, aggregated to weekly bins. SFW subreddits (treatment, blue) are surfaced in AI Overviews; NSFW subreddits (control, red) are excluded from AI Overviews. The gray dashed line marks May 14, 2024, when Google launched AI Overviews for all U.S.\ users. The red dashed line marks August 15, 2024, the date of the international rollout.}
\end{figure}

Figure~\ref{fig:model_free} plots the average daily comments and comment authors for SFW and NSFW subreddits from January 2024 to July 2025. The gray dashed line marks May 14, 2024, when Google launched AI Overviews for all U.S.\ users, and the red dashed line marks August 15, 2024, the date of the international rollout. For both outcomes, the two groups evolve in parallel before the treatment date, consistent with the parallel trends assumption. After the rollout, SFW subreddits show a sustained increase in both comments and comment authors, whereas NSFW subreddits remain flat or decline slightly, indicating a widening engagement gap between the two groups.

Table~\ref{tab:summary_stats} reports summary statistics for our two primary outcomes by treatment group. SFW and NSFW subreddits differ substantially in baseline activity: SFW subreddits tend to be larger, more discussion-oriented communities, while many NSFW subreddits are media-sharing communities with less textual engagement. Consistent with the model-free evidence in Figure~\ref{fig:model_free}, the pre- and post-treatment means suggest differential trends: SFW subreddits show increases in daily comments and comment authors from the pre- to post-period, while NSFW subreddits show small declines in both outcomes. These raw differences motivate our main analysis in the following section.

\begin{table}[t!]
\centering
\caption{Summary Statistics}
\label{tab:summary_stats}
\renewcommand{\arraystretch}{1.15}
\scriptsize
\begin{threeparttable}
\begin{tabular}{ll|c|rrrr|rr|rr}
\hline\hline
 & & & \multicolumn{4}{c|}{All observations} & \multicolumn{2}{c|}{Before treatment} & \multicolumn{2}{c}{After treatment} \\
Daily outcome & Group & Subreddits & Mean & Std.\ dev & Max & Min & Mean & Std.\ dev & Mean & Std.\ dev \\
\hline
\multirow{2}{*}{Comments/Day}
 & SFW  & \num{70683} & 110.94 & 822.34 & \num{628964} & 0 & 105.36 & 817.76 & 114.54 & 825.27 \\
 & NSFW & \num{34329} &  30.18 & 269.64 &  \num{39497} & 0 &  32.27 & 295.40 &  28.83 & 251.57 \\[4pt]
\multirow{2}{*}{Comment Authors/Day}
 & SFW  & \num{70683} & 57.51 & 380.20 & \num{129373} & 0 & 53.64 & 370.98 & 60.01 & 386.02 \\
 & NSFW & \num{34329} & 12.77 &  73.46 &  \num{11780} & 0 & 12.95 &  73.97 & 12.66 &  73.12 \\
\hline\hline
\end{tabular}
\begin{tablenotes}
\scriptsize
\item \textit{Note}: Each observation is a subreddit-day. The panel is perfectly balanced: \num{70683} SFW and \num{34329} NSFW subreddits $\times$ 578 calendar days (January 1, 2024 to July 31, 2025), yielding \num{60696936} total observations. The pre-treatment period runs from January 1 to August 14, 2024 (227 days) and the post-treatment period from August 15, 2024 to July 31, 2025 (351 days). Cohort defined as subreddits with $\geq 10$ submissions in December 2023, excluding user-profile pages, quarantined subreddits, and non-public (private or restricted) subreddits. NSFW classification locked from December 2023.
\end{tablenotes}
\end{threeparttable}
\end{table}


\section{Empirical Analysis}\label{sec:empirical}

In this section, we report the effect of Google AI Overviews on Reddit comment engagement. We first describe our identification strategy, then present the main analysis results.

\subsection{Model Specification}\label{sec:identification}

Our identification strategy leverages two facts: (1) the introduction of Google AI Overviews provides an exogenous shock to the Reddit platform; and (2) Google's content moderation policy ensures that both SFW and NSFW communities are indexed in organic search results, but only SFW communities are referenced in AI Overviews while NSFW communities are excluded.
We estimate the causal effect of AI Overviews on Reddit's user engagement using a difference-in-differences (DiD) design that compares changes in activity for SFW subreddits (treated) to changes for NSFW subreddits (control) around the AI Overviews rollout. The model specification is:
\begin{equation}\label{eq:did_standard}
    Y_{it} = \beta \times \text{SFW}_i \times \text{Post}_t + \alpha_i + \gamma_t + \varepsilon_{it},
\end{equation}
where $i$ denotes subreddit, $t$ denotes day, and $Y_{it}$ is an activity outcome (e.g., the number of comments or unique comment authors) for subreddit $i$ on day $t$. $\text{SFW}_i$ equals 1 if subreddit $i$ is classified as SFW and 0 if NSFW. $\text{Post}_t$ equals 1 if day $t$ falls on or after August 15, 2024, and 0 otherwise. The subreddit fixed effect $\alpha_i$ captures time-invariant characteristics of each subreddit (e.g., baseline popularity, community size), absorbing the main effect of $\text{SFW}_i$. The day fixed effect $\gamma_t$ controls for platform-wide time trends, absorbing the main effect of $\text{Post}_t$. $\varepsilon_{it}$ is the error term. The coefficient $\beta$ is our parameter of interest, capturing the differential change in activity for SFW relative to NSFW subreddits after the introduction of AI Overviews.




\subsection{Main Results}\label{sec:main_results}

Table~\ref{tab:main_did} reports the DiD estimates from Equation~\eqref{eq:did_standard}. Daily comments per SFW subreddit significantly increased by \num{12.621} relative to NSFW subreddits ($p < 0.001$), a 12.0\% increase relative to the SFW pre-treatment mean of 105.4 comments per day. Daily comment authors significantly increased by \num{6.667} ($p < 0.001$), a 12.4\% increase relative to the pre-treatment mean of 53.6 comment authors per day. Both outcomes pass the pre-trend tests in Section~\ref{sec:robustness}, which supports a causal interpretation of these effects.

These effects indicate that AI Overviews substantially increase user engagement in communities whose content is surfaced in AI Overviews. Comments grew by 12.0\% while comment authors grew by 12.4\%, indicating that the increase is driven primarily by the extensive margin: it comes from more people participating in discussions rather than from existing commenters writing more. If the effect operated through the intensive margin, we would expect comments to grow substantially faster than comment authors, but the two rates are nearly identical.

\begin{table}[t!]
\centering
\caption{Effect of Google AI Overviews on Reddit Comment Engagement}
\label{tab:main_did}
\renewcommand{\arraystretch}{1.15}
\setlength{\tabcolsep}{10pt}
\scriptsize
\begin{threeparttable}
\begin{tabular}{l|cc}
\hline\hline
 & Comments/Day & Comment Authors/Day \\
 & (1) & (2) \\
\hline
$\text{SFW} \times \text{Post}$ & \num{12.621}$^{***}$ & \num{6.667}$^{***}$ \\
 & (\num{1.225}) & (\num{0.496}) \\[4pt]
\hline
Effect size & 12.0\% & 12.4\% \\
Observations & \num{60696936} & \num{60696936} \\
\hline\hline
\end{tabular}
\begin{tablenotes}
\scriptsize
\item \textit{Note}: This table reports DiD estimates from Equation~\eqref{eq:did_standard}. Each column is a separate regression with the indicated outcome variable. The sample is a balanced panel of \num{105012} subreddits $\times$ 578 days (January 1, 2024 -- July 31, 2025). Treatment date is August 15, 2024. Standard errors clustered at the subreddit level are reported in parentheses. Effect size computed as coefficient divided by pre-treatment SFW mean. $^{***}p < 0.001$; $^{**}p < 0.01$; $^{*}p < 0.05$; $^{\dagger}p < 0.1$.
\end{tablenotes}
\end{threeparttable}
\end{table}

\subsection{Effect Distribution}\label{sec:quantile}

To assess whether the positive effect is broad-based or concentrated among a few highly active subreddits, we estimate quantile treatment effects using the Changes-in-Changes (CiC) estimator of \citet{athey2006}. CiC generalizes DiD from means to the full outcome distribution: at each quantile $\tau$, it estimates the quantile treatment effect on the treated (QTET) as the difference between the observed $\tau$-th quantile of SFW outcomes and the counterfactual $\tau$-th quantile SFW subreddits would have realized absent AI Overviews. We collapse the panel to pre- and post-treatment means per subreddit and estimate QTETs at deciles Q10 through Q90 for both comments per day and comment authors per day. Table~\ref{tab:cic} reports each QTET alongside the SFW pre-treatment baseline and the implied effect size.

\begin{table}[t!]
\centering
\caption{Changes-in-Changes: Quantile Treatment Effects on the Treated}
\label{tab:cic}
\renewcommand{\arraystretch}{1.1}
\setlength{\tabcolsep}{10pt}
\scriptsize
\begin{threeparttable}
\begin{tabular}{l|rrr|rrr}
\hline\hline
 & \multicolumn{3}{c|}{Comments/Day} & \multicolumn{3}{c}{Comment Authors/Day} \\
 & Base & QTET & Effect size & Base & QTET & Effect size \\
\hline
Q10 & 0.00 & 0.000 & -- & 0.00 & 0.000 & -- \\
Q20 & 0.06 & 0.006$^{***}$ & 9.9\% & 0.04 & 0.006$^{***}$ & 12.9\% \\
Q30 & 0.44 & 0.171$^{***}$ & 39.2\% & 0.33 & 0.134$^{***}$ & 40.0\% \\
Q40 & 1.39 & 0.895$^{***}$ & 64.3\% & 1.00 & 0.667$^{***}$ & 66.7\% \\
Q50 & 3.27 & 2.074$^{***}$ & 63.4\% & 2.27 & 1.393$^{***}$ & 61.4\% \\
Q60 & 7.46 & 3.738$^{***}$ & 50.1\% & 4.96 & 2.313$^{***}$ & 46.7\% \\
Q70 & 17.64 & 6.048$^{***}$ & 34.3\% & 11.35 & 2.969$^{***}$ & 26.2\% \\
Q80 & 48.39 & 9.356$^{***}$ & 19.3\% & 29.20 & 3.499$^{***}$ & 12.0\% \\
Q90 & 171.73 & 35.256$^{***}$ & 20.5\% & 95.98 & 15.712$^{***}$ & 16.4\% \\
\hline\hline
\end{tabular}
\begin{tablenotes}
\scriptsize
\item \textit{Note}: This table reports CiC estimates \citep{athey2006} for the two outcomes in Table~\ref{tab:main_did}, both of which pass the pre-trend tests in Section~\ref{sec:parallel_trends}. ``Base'' is the pre-treatment level of the treated group at each quantile (average daily outcome). ``QTET'' is the quantile treatment effect on the treated. ``Effect size'' is the QTET as a percentage of the baseline. $^{***}p < 0.001$; $^{**}p < 0.01$; $^{*}p < 0.05$; $^{\dagger}p < 0.1$.
\end{tablenotes}
\end{threeparttable}
\end{table}

All QTETs from Q20 to Q90 are positive and significant at the 0.1\% level for both outcomes; Q10 has zero baseline activity, where the estimator is uninformative. The relative effects peak at Q40 (64\% for comments and 67\% for comment authors), and are smaller at the bottom of the distribution (10\% and 13\% at Q20) and at the top (21\% and 16\% at Q90). Absolute QTETs, by contrast, rise monotonically with the quantile, from 0.01 additional comments per day at Q20 to 35.3 at Q90. Mid-sized communities thus benefit proportionally more from AI Overviews, while large communities account for most of the absolute increase in engagement.

This pattern shows that the largest proportional gains are concentrated in communities in the mid-distribution that were neither inactive nor dominant before treatment, indicating that AI Overviews redirect attention to communities users would have been less likely to find through traditional ranked search or Reddit's own recommendation system.

\subsection{Effect Across Subreddit Size}\label{sec:size_heterogeneity}

We further investigate whether AI Overviews benefit communities of all sizes or primarily amplifies already-large ones. To examine this, we split subreddits into quartiles by December 2023 subscriber count and re-estimate Equation~\eqref{eq:did_standard} within each quartile.

Table~\ref{tab:size_heterogeneity} reports the results. The treatment effect is positive across all four size quartiles for both outcomes, and statistically significant at the 0.1\% level for Q2 through Q4, at the 1\% level for comment authors in Q1, and at the 5\% level for comments in Q1. The absolute magnitude increases monotonically with subreddit size: the largest subreddits (Q4) gain approximately 35.3 additional comments per day, compared to 0.7 for the smallest (Q1). The relative effect, by contrast, peaks in the small-to-medium quartile (Q2), where comments increase by 56\% and comment authors by 77\% over their pre-treatment baselines, and declines toward both tails (49\% and 61\% at Q1; 10\% and 9\% at Q4). 

The substantive implication mirrors the results in Section~\ref{sec:quantile}. Small and mid-sized communities receive the largest proportional gains. By citing these communities directly in generated answers, AI Overviews surface ``hidden gems'' that users would have been less likely to discover through organic results or Reddit's own homepage feed.

\begin{table}[t!]
\centering
\caption{Effect of AI Overviews Across Subreddit Size Quartiles}
\label{tab:size_heterogeneity}
\renewcommand{\arraystretch}{1.1}
\setlength{\tabcolsep}{4pt}
\scriptsize
\begin{threeparttable}
\begin{tabular}{ll|cc|cc}
\hline\hline
 & & \multicolumn{2}{c|}{Comments/Day} & \multicolumn{2}{c}{Comment Authors/Day} \\
Quartile & Subscribers & $\hat{\beta}$ & Effect size & $\hat{\beta}$ & Effect size \\
\hline
Q1 (Smallest) & 0--238 & \num{0.731}$^{*}$ & 48.8\% & \num{0.407}$^{**}$ & 61.3\% \\
 & & (\num{0.372}) & & (\num{0.132}) & \\
Q2 & 239--\num{2080} & \num{6.096}$^{***}$ & 55.6\% & \num{3.923}$^{***}$ & 77.2\% \\
 & & (\num{1.749}) & & (\num{1.024}) & \\
Q3 & \num{2081}--\num{12385} & \num{11.079}$^{***}$ & 35.7\% & \num{6.791}$^{***}$ & 41.8\% \\
 & & (\num{1.613}) & & (\num{0.669}) & \\
Q4 (Largest) & $>$\num{12385} & \num{35.326}$^{***}$ & 10.2\% & \num{15.181}$^{***}$ & 8.6\% \\
 & & (\num{4.624}) & & (\num{1.600}) & \\
\hline\hline
\end{tabular}
\begin{tablenotes}
\scriptsize
\item \textit{Note}: Each row reports DiD estimates from Equation~\eqref{eq:did_standard} estimated on the subsample of subreddits in the indicated size quartile. Quartiles are defined by each subreddit's maximum subscriber count in December 2023, pooled across SFW and NSFW subreddits. Each quartile contains approximately \num{26253} subreddits. All specifications include subreddit and day fixed effects with standard errors clustered at the subreddit level (in parentheses). ``Effect size'' denotes the coefficient divided by the pre-treatment SFW mean. $^{***}p < 0.001$; $^{**}p < 0.01$; $^{*}p < 0.05$; $^{\dagger}p < 0.1$.
\end{tablenotes}
\end{threeparttable}
\end{table}





\section{Mechanism}\label{sec:mechanism}

To understand why our results show that AI search increases user engagement while prior studies document substitution effects on platform engagement, we examine how AI Overviews affect different types of content in this section.

\subsection{Search Goods and Experience Goods}\label{sec:theory}

To classify content type across the online content ecosystem, we draw on the classic distinction between search goods and experience goods \citep{nelson1970}. Search goods are products whose quality can be assessed before consumption; in the context of online content, these are communities that primarily host factual, verifiable information such as news, sports scores, and reference material. Experience goods, by contrast, require consumption to evaluate; on Reddit, these are communities centered on opinions, discussions, personal advice, and subjective experiences.

This distinction generates two opposing forces. For search goods, AI summaries can directly answer users' queries and substitute for the underlying content, attenuating engagement. For experience goods, AI cannot replicate authentic human discussion; instead, by surfacing Reddit prominently in search results, AI Overviews channel new users toward these communities. Our hypothesis is therefore that experience-good communities exhibit larger positive treatment effects than search-good communities, as the discovery effect dominates where AI cannot substitute for human interaction.


We classify each subreddit as a search good or an experience good using Google's Gemini 3 Flash large language model. The model receives the subreddit's name, title, and public description (drawn from the Arctic Shift Reddit Subreddits Metadata dump \citep{heitmann2024arcticshift}, distributed via Academic Torrents) and applies a single decision rule adapting the search-versus-experience distinction of \citet{nelson1970} to online content: if a user could obtain most of the subreddit's value by reading posts alone, with no comments or interaction, the subreddit is a search good; otherwise it is an experience good. SFW and NSFW subreddits are evaluated under identical criteria, and the prompt makes no reference to AI, substitution, or post-treatment outcomes, keeping the classification independent of the engagement metrics we study. The model returns a binary label, a confidence score on a 0.0--1.0 scale (with low scores reserved for uninformative names and descriptions), and a brief justification.

Of the \num{105012} subreddits in our cohort, the model successfully classifies \num{104975}: \num{59450} (56.6\%) as search goods and \num{45525} (43.4\%) as experience goods. The remaining 37 subreddits were either blocked by Gemini's content-safety filter or returned responses that could not be parsed as either label.  \hyperref[sec:appendix_classification]{Appendix~B} provides the full classification prompt and the SFW/NSFW breakdown.

\subsection{Results}\label{sec:mechanism_results}

We test the search--experience prediction using a triple-difference model that pools all subreddits and directly estimates whether treatment effects differ by content type:
\begin{equation}\label{eq:triple_diff}
Y_{it} = \beta_1 (\text{SFW}_i \times \text{Post}_t) + \beta_2 (\text{Exp}_i \times \text{Post}_t) + \beta_3 (\text{SFW}_i \times \text{Post}_t \times \text{Exp}_i) + \alpha_i + \gamma_t + \varepsilon_{it}
\end{equation}
where $\text{Exp}_i$ is an indicator for experience-based subreddits. The subreddit fixed effects $\alpha_i$ absorb all time-invariant subreddit characteristics, including the main effects of $\text{SFW}_i$, $\text{Exp}_i$, and their interaction; the day fixed effects $\gamma_t$ absorb $\text{Post}_t$. The coefficient $\beta_1$ captures the DiD treatment effect for search goods (the reference category), $\beta_2$ captures differential post-treatment trends between experience and search goods in the NSFW control group, and $\beta_3$ is the key parameter: the \textit{additional} treatment effect for experience goods relative to search goods.

Table~\ref{tab:triple_diff} reports the results for comments and comment authors per day. The triple-difference coefficient $\beta_3$ is positive and highly significant for both outcomes: $6.40$ for comment authors ($p < 0.001$) and $9.87$ for comments ($p < 0.001$). For comment authors, the implied treatment effect for experience goods ($\beta_1 + \beta_3 = 9.93$) is 2.8 times the search-good effect ($\beta_1 = 3.54$); for comments, the ratio is 2.3 ($17.56$ vs.\ $7.68$). 

We find that, given the forum nature of Reddit, search-good and experience-good communities both show significant positive effects, indicating that the discovery channel operates across all content types. However, the effect is substantially amplified where AI cannot substitute for human interaction. The results are robust to restricting the sample to the 81\% of subreddits that receive high-confidence ($\geq 0.8$) classifications from Gemini (Appendix~\ref{sec:appendix_highconf}).

\begin{table}[t!]
\centering
\caption{Triple-Difference: Formal Test of Search vs Experience Heterogeneity}
\label{tab:triple_diff}
\renewcommand{\arraystretch}{1.1}
\setlength{\tabcolsep}{8pt}
\scriptsize
\begin{threeparttable}
\begin{tabular}{l|cc}
\hline\hline
& Comments/Day & Comment Authors/Day \\
\hline
SFW $\times$ Post & $7.68^{***}$ & $3.54^{***}$ \\
& $(1.400)$ & $(0.593)$ \\[4pt]
Exp $\times$ Post & $-0.72$ & $-0.72^{\dagger}$ \\
& $(1.662)$ & $(0.383)$ \\[4pt]
SFW $\times$ Post $\times$ Exp & $9.87^{***}$ & $6.40^{***}$ \\
& $(2.546)$ & $(0.994)$ \\[4pt]
\hline
Implied experience DiD ($\beta_1 + \beta_3$) & $17.56$ & $9.93$ \\
Experience / Search ratio & $2.3\times$ & $2.8\times$ \\
\hline
Subreddit FE & Yes & Yes \\
Day FE & Yes & Yes \\
Subreddits & \num{104975} & \num{104975} \\
Observations & \num{60675550} & \num{60675550} \\
\hline\hline
\end{tabular}
\begin{tablenotes}
\scriptsize
\item \textit{Note}: Each column reports estimates from Equation~\eqref{eq:triple_diff}. $\beta_1$ is the DiD treatment effect for search goods. $\beta_3$ is the additional treatment effect for experience goods relative to search goods. Standard errors clustered at the subreddit level in parentheses. $^{***}p < 0.001$; $^{**}p < 0.01$; $^{*}p < 0.05$; $^{\dagger}p < 0.1$.
\end{tablenotes}
\end{threeparttable}
\end{table}

\section{From Static Summaries to Conversational Search}\label{sec:aimode}





\begin{figure}[t!]
\centering
\caption{Google AI Mode: Conversational Search Interface}
\includegraphics[width=0.85\textwidth]{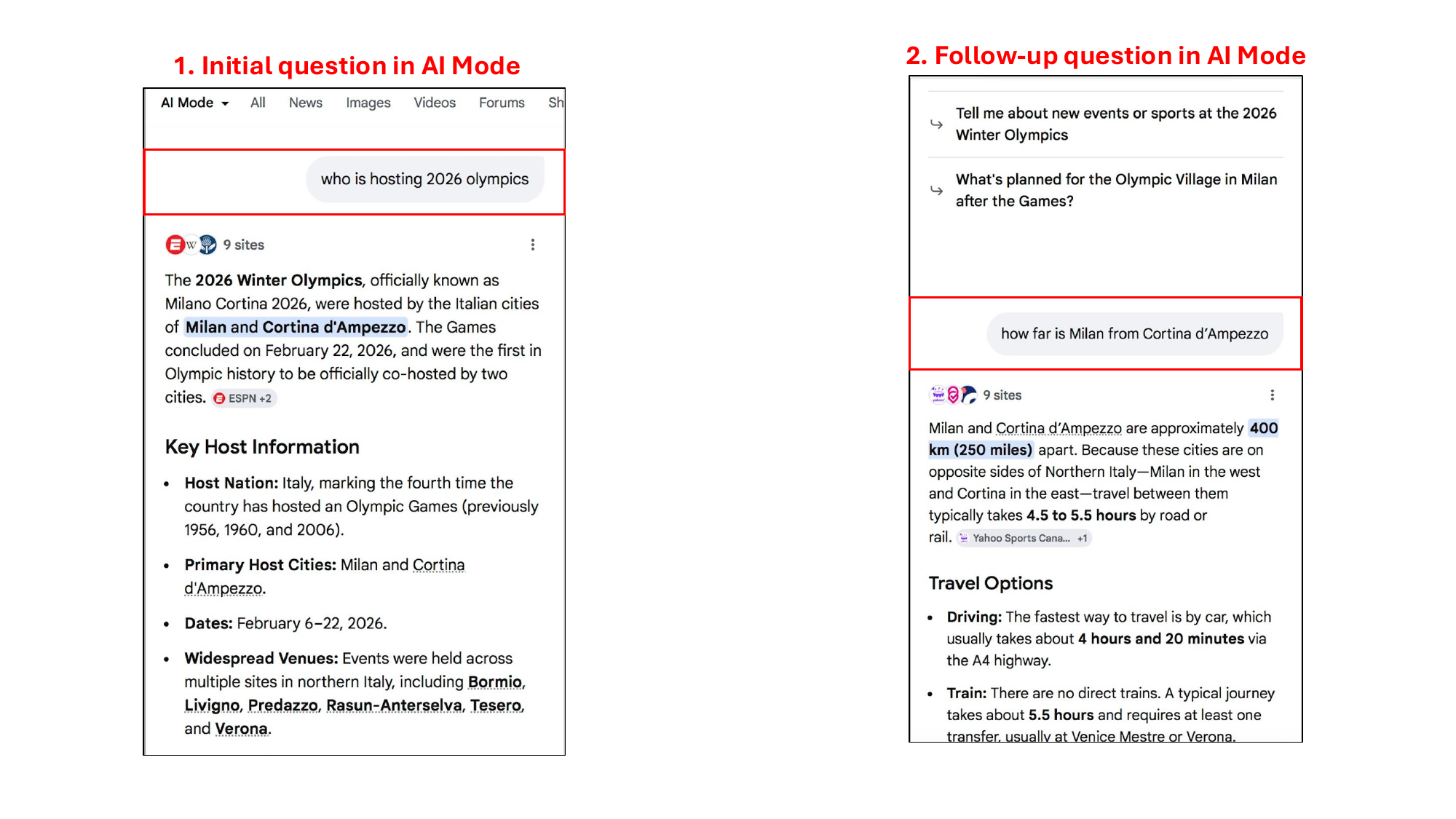}
\label{fig:ai_mode}
\end{figure}

How does the experience-good premium change when AI search becomes more interactive? A more interactive interface could satisfy users' need for richer detail directly, weakening their incentive to visit the source. Google AI Mode provides exactly this setting: it transforms AI search from static summaries into a conversational interface where users can chat with search results. Figure~\ref{fig:ai_mode} illustrates this conversational format, showing how users can pose follow-up questions within the AI Mode interface.

AI Mode rolled out globally on August 21, 2025 (with early access in the United States beginning in May 2025). AI Mode follows the same content moderation policy as AI Overviews and does not reference NSFW content in its responses \citep{google2025aio,google2025content}, preserving the SFW-NSFW treatment-control contrast under the conversational interface. We extend our panel through December 2025 to test whether the discovery effect persists under this more interactive format.

We augment the triple-difference specification from Equation~\eqref{eq:triple_diff} by adding a second set of interaction terms that activate after the AI Mode launch:
\begin{align}\label{eq:aimode}
Y_{it} = 
& \beta_1 (\text{SFW}_i \times \text{Post}_t) + \beta_2 (\text{Exp}_i \times \text{Post}_t) + \beta_3 (\text{SFW}_i \times \text{Post}_t \times \text{Exp}_i) \nonumber \\
&+ \beta_4 (\text{SFW}_i \times \text{Post}^{\text{Mode}}_t) + \beta_5 (\text{Exp}_i \times \text{Post}^{\text{Mode}}_t) + \beta_6 (\text{SFW}_i \times \text{Post}^{\text{Mode}}_t \times \text{Exp}_i) + \alpha_i + \gamma_t +  \varepsilon_{it}
\end{align}
where $\text{Post}_t$ equals 1 for all dates from August 15, 2024 onward, $\text{Post}^{\text{Mode}}_t$ equals 1 for all dates from August 21, 2025 onward, and all other variables are defined as before. We use the August 21, 2025 global rollout as the AI Mode treatment date, when AI Mode became available in countries that account for the majority of Reddit's users. The coefficients $\beta_1$ through $\beta_3$ capture the baseline treatment effects of AI Overviews, while $\beta_4$ through $\beta_6$ capture the incremental shifts after AI Mode. The key parameter is $\beta_6$: the change in the experience-good premium after AI Mode. The experience-good premium is $\beta_3$ before AI Mode and $\beta_3 + \beta_6$ after. If AI Mode attenuates the discovery effect, $\beta_6$ should be negative.

Table~\ref{tab:aimode} reports the results. The experience-good premium under AI Overviews is large and highly significant: $6.29$ additional comment authors per day ($p < 0.001$) and $9.54$ additional comments per day ($p < 0.001$), consistent with our earlier triple-difference findings. The key finding is that the experience-good premium is substantially attenuated after AI Mode. For comments, the premium falls by $9.71$ ($p < 0.001$), from $9.54$ to $-0.17$. For comment authors, it falls by $3.72$ ($p < 0.001$), from $6.29$ to $2.57$, a 59\% decline. The base treatment effect for search-good communities ($\beta_4$) also declines after AI Mode ($-2.91$, $p = 0.029$ for comments; $-1.75$, $p = 0.010$ for comment authors), but the additional drop concentrated on experience goods ($\beta_6$) is more than twice as large, indicating that the attenuation is driven primarily by the experience-good premium documented in Section~\ref{sec:mechanism}.

One might worry that this attenuation reflects not the interface change but a gradual maturation of AI search that raised user trust and reduced click-through over time. Our design speaks against this in three ways. First, a trust-maturation channel would predict attenuation concentrated among search goods: for a search good, the summary states the fact the user wants, so trusting it removes the click, whereas for an experience good, the user wants the discussion itself, so trusting the summary gives no reason to skip the visit. Our results in Table~\ref{tab:aimode} show the reverse: the experience-good premium attenuates proportionally more ($\beta_6$) than the search-good effect does ($\beta_4$). The decline lands precisely where a substitution-by-trust story predicts it should not, and where a substitution-by-interface story predicts it should. Second, a gradual rise in user trust would erode the positive effect on user engagement slowly over time, yet the monthly event study in Appendix~\ref{sec:appendix_event_study} (Figure~\ref{fig:event_study_main}) shows the effect holding steady from April through July 2025 and declining only after the AI Mode launch in August 2025. A discrete break at the interface change, rather than a slow drift, is inconsistent with gradual trust maturation. Third, the day fixed effects absorb any trend in AI trust common across communities. Together, this evidence supports the interface change as the primary driver of the decline in user engagement.


\begin{table}[t!]
\centering
\caption{AI Mode Attenuation of the Experience-Good Premium}
\label{tab:aimode}
\renewcommand{\arraystretch}{1.1}
\setlength{\tabcolsep}{8pt}
\scriptsize
\begin{threeparttable}
\begin{tabular}{l|cc}
\hline\hline
& Comments/Day & Comment Authors/Day \\
\hline
\multicolumn{3}{l}{\textit{Baseline post-AI-Overviews effects}} \\[2pt]
\quad SFW $\times$ Post & $7.67^{***}$ & $3.53^{***}$ \\
                        & $(1.41)$ & $(0.60)$ \\[4pt]
\quad Exp $\times$ Post & $-0.70$ & $-0.73^{\dagger}$ \\
                        & $(1.69)$ & $(0.39)$ \\[4pt]
\quad SFW $\times$ Post $\times$ Exp & $9.54^{***}$ & $6.29^{***}$ \\
                        & $(2.57)$ & $(1.00)$ \\[4pt]
\hline
\multicolumn{3}{l}{\textit{Incremental post-AI-Mode shifts}} \\[2pt]
\quad SFW $\times$ Post$^{\text{Mode}}$ & $-2.91^{*}$ & $-1.75^{**}$ \\
                        & $(1.33)$ & $(0.68)$ \\[4pt]
\quad Exp $\times$ Post$^{\text{Mode}}$ & $0.47$ & $-0.03$ \\
                        & $(1.19)$ & $(0.30)$ \\[4pt]
\quad SFW $\times$ Post$^{\text{Mode}}$ $\times$ Exp & $-9.71^{***}$ & $-3.72^{***}$ \\
                        & $(2.35)$ & $(0.97)$ \\[4pt]
\hline
Exp premium (post-AI Overviews) & $9.54$ & $6.29$ \\
Exp premium (post-AI Mode) & $-0.17$ & $2.57$ \\
Attenuation & $102\%$ & $59\%$ \\
\hline
Subreddit FE & Yes & Yes \\
Day FE & Yes & Yes \\
Subreddits & \num{104975} & \num{104975} \\
Observations & \num{76736725} & \num{76736725} \\

\hline\hline
\end{tabular}
\begin{tablenotes}
\scriptsize
\item \textit{Note}: Each column reports estimates from Equation~\eqref{eq:aimode}. The panel is extended through December 2025. $\text{Post}$ equals one for all dates from August 15, 2024 onward; $\text{Post}^{\text{Mode}}$ equals one for all dates from August 21, 2025 onward (global AI Mode launch). The first three rows are baseline post-treatment effects; the next three rows are incremental shifts after AI Mode. The experience-good premium is SFW $\times$ Post $\times$ Exp before AI Mode, and the sum of that coefficient and SFW $\times$ Post$^{\text{Mode}}$ $\times$ Exp after. Standard errors clustered at the subreddit level appear in parentheses. $^{***}p < 0.001$; $^{**}p < 0.01$; $^{*}p < 0.05$; $^{\dagger}p < 0.1$.
\end{tablenotes}
\end{threeparttable}
\end{table}

These results carry two implications. First, the discovery effect for experience-based communities is driven primarily by the limitations of static summaries: a conversational interface that delivers richer detail on the search page itself satisfies the demand that previously sent users to Reddit. Second, the attenuation differs across margins. For comments, the experience-good premium turns slightly negative after AI Mode ($-0.17$), indicating that the depth of redirected engagement is fully absorbed by the conversational interface. For comment authors, the premium remains positive ($2.57$), suggesting that platform-native features such as community discussion and peer feedback retain value at the participation margin that a more interactive AI interface cannot fully replicate. Together, these findings indicate that the impact of AI search on the content ecosystem depends not only on content type but also on the AI search interface design.


\section{Robustness Tests}\label{sec:robustness}

We conduct several analyses to assess the robustness of our main findings and validate the identifying assumptions. Detailed results are reported in \hyperref[sec:appendix_robustness]{Appendix~A}.

\subsection{Parallel Trends}\label{sec:parallel_trends}

The validity of our DiD design rests on the assumption that SFW and NSFW subreddits would have followed parallel activity trajectories in the absence of AI Overviews. We assess this assumption using three complementary approaches.

\noindent \textbf{Event Study.} We estimate a monthly event study that allows the treatment effect to vary by period (see \hyperref[sec:appendix_event_study]{Appendix~A.1} for the full specification and Figure~\ref{fig:event_study_main}). For both outcomes, the pre-treatment coefficients are all negative relative to the July 2024 reference. For comments, they are statistically insignificant from January through April 2024 and significantly negative in May and June; for comment authors, they are insignificant in January and February and significantly negative from March through June, consistent with the negative slope in the restricted pre-window below. Both series then rise toward the reference after the U.S.\ AI Overviews launch in May 2024, a pattern we read as early treatment exposure rather than a divergent pre-trend. The slope test below detects no significant linear pre-trend, and we conclude that our reported effect is a conservative lower bound on the true effect.

\noindent \textbf{Pre-Trend Slope Test.} We test for a differential pre-trend by interacting the SFW indicator with a linear time index. Over the full pre-treatment window (January--July 2024), the slope is insignificant for both comments/day and comment authors/day. Restricting to January--April 2024, the window before AI Overviews (AIO) launched in any country, leaves the comments slope insignificant; the comment-authors slope is negative ($p = 0.026$), a pre-trend that works against our positive estimate rather than inflating it (see Appendix~\ref{sec:appendix_pretrend}).

\noindent \textbf{Sensitivity Analysis.} We further assess the sensitivity of our results using the HonestDiD framework of \citet{rambachan2023}, which asks: how large would post-treatment trend violations need to be, relative to the maximum pre-treatment shift, to render the treatment effect statistically insignificant? The breakdown value $\bar{M}$ summarizes the result, with higher values indicating greater robustness. For the average treatment effect across all post-treatment periods, comments/day is robust up to $\bar{M} = 0.30$ and comment authors/day up to $\bar{M} = 0.50$, indicating that the main effects remain significant as long as post-treatment violations of parallel trends are no larger than 30\% (comments) and 50\% (comment authors) of the largest pre-treatment violation. Full sensitivity plots are reported in \hyperref[sec:appendix_honestdid]{Appendix~A.3}.

\subsection{Treatment Date Validation}\label{sec:treatment_date}

We assess the sensitivity of our DiD estimate to the specification of the treatment date in two ways.

\noindent \textbf{Alternative Treatment Dates.} AI Overviews launched in the U.S. on May 14, 2024, three months before our main August 15 treatment date (the international expansion). We re-estimate the DiD using May 14 as the treatment date. Point estimates are positive and statistically significant under either date, with the August magnitudes somewhat larger, consistent with the international rollout adding an incremental effect on top of the U.S.\ launch (see \hyperref[sec:appendix_treatment_date]{Appendix~A.4}).

\noindent \textbf{Placebo Test at Pre-AIO Policy Dates.} We test for SFW-NSFW divergence around three pre-AIO events that might confound our results: the Reddit-Google content licensing deal, the Google March 2024 Core Update, and the Reddit IPO. Re-estimating the DiD using each as a fake treatment date over a 30-day window before and after the event yields statistically insignificant placebo estimates for comments/day at all three dates, and insignificant placebo estimates for comment authors/day at two of the three dates. The only significant placebo estimate (comment authors at the Reddit IPO date) is negative, meaning SFW comment-author activity was already trending below NSFW before AIO; this makes our reported positive DiD effect a conservative lower bound on the true effect (see \hyperref[sec:appendix_placebo_policy]{Appendix~A.5}).

\subsection{Control Group Validity}\label{sec:control_validity}

A residual concern is whether NSFW activity is itself in secular decline due to platform or regulatory pressure. The model-free evidence in Figure~\ref{fig:model_free} shows that NSFW activity remains roughly flat throughout the post-treatment period. Moreover, the recent wave of U.S. state age-verification laws does not legally apply to Reddit because the platform's overall content mix falls below the one-third threshold for sexual material harmful to minors that triggers these statutes (see \hyperref[sec:appendix_nsfw_content]{Appendix~A.6}).

\subsection{Alternative Cohort Definitions}\label{sec:alt_cohort}

We replicate the analysis using alternative cohort activity thresholds of 20 and 50 minimum submissions in December 2023, yielding cohorts of \num{73410} and \num{46096} subreddits, respectively. The results are quantitatively similar across all three thresholds: the relative effect on comments/day is 11.3\% (min 20) and 9.5\% (min 50), and the relative effect on comment authors/day is 11.9\% (min 20) and 10.0\% (min 50), close to the primary estimates of 12.0\% and 12.4\% from the min 10 cohort (see Appendix~\ref{sec:appendix_cohort}).

\subsection{Influential Periods and Outliers}\label{sec:influential}

We confirm that the main effect is not driven by a single high-activity month or by a small number of extreme observations. Dropping the peak month of February 2025 leaves the estimates essentially unchanged (relative effects of 11.2\% for comments/day and 11.8\% for comment authors/day). Winsorizing both outcomes at the 1st and 99th percentiles yields relative effects of 12.1\% and 12.0\%, nearly identical to the main estimates (see Appendix~\ref{sec:appendix_influential}).


\section{Discussion and Conclusion}\label{sec:discussion}

In this paper, we study how AI search affects the online content ecosystem. Using a difference-in-differences design that compares SFW and NSFW subreddits on Reddit, we find that Google AI Overviews cause a statistically significant increase in comment engagement: daily comments rise by 12.0\% and daily comment authors by 12.4\%, with both outcomes passing pre-trend tests. These findings challenge the substitution hypothesis that has emerged from studies of factual-content platforms such as Wikipedia \citep{khosravi2026}. 

To reconcile these contrasting results, we develop a search-experience framework that distinguishes when AI search complements versus substitutes for a platform's content; we draw on \citet{nelson1970} to classify content into the two types: search-good content and experience-good content. For search-good content, where information is factual and AI-summarizable, AI Overviews can directly satisfy users' needs and partially substitute for visits to the source platform. For experience-good content, where value derives from personal perspectives, subjective advice, and interactive discussion, AI summaries cannot replace the underlying content; instead, they serve as a discovery channel that directs users to it. Within Reddit, both content types show positive treatment effects, but experience-good communities respond 2.3 to 2.8 times more strongly than search-good communities, consistent with the discovery channel dominating where AI cannot substitute for human interaction. Across platforms, the framework correctly predicts the directional contrast between Reddit (positive) and Wikipedia (negative). We further exploit the subsequent launch of Google AI Mode to show that the experience-good premium is largely attenuated when the AI search interface shifts from static summaries to a conversational format, falling by 59--102\%, consistent with the framework's prediction when AI begins to substitute for the interactivity that previously drove users to the platform.

Our results have three main implications. First, for content platforms, the effect of AI search depends critically on the type of content a platform hosts. Platforms centered on discussion, opinions, and subjective experiences may benefit from AI search, as AI summaries serve as advertisements for richer content that must be consumed in its original form. Factual-content platforms, by contrast, face substitution risk, as AI can directly satisfy the information need. The ``AI kills content platforms'' narrative is thus accurate for some platforms but incomplete as a general characterization. Platform operators should assess the composition of content types within their ecosystem: those with rich discussion content may benefit from ensuring discoverability by AI systems, while those with primarily factual content may need to differentiate through interactive features and community engagement that AI cannot replicate.

Second, for search engine providers, our AI Mode results highlight a design trade-off in how AI is integrated into search. Static AI summaries (AI Overviews) act as a discovery channel that drives users to experience-good platforms, creating a complementary relationship. Conversational AI interfaces (AI Mode), however, begin to substitute for the interactive discussions that previously required visiting the platform. As search engines continue to develop more interactive AI features, this trade-off between driving engagement to content platforms and retaining users within the search interface will become increasingly consequential for the health of the broader content ecosystem.

Third, for regulators, our findings counsel against one-size-fits-all rules. The ongoing lawsuits and antitrust scrutiny of AI search presume a uniform harm to content platforms, yet the effect we document ranges from substitution to complementarity depending on the type of content a platform hosts and the interface a search engine adopts. Remedies calibrated to factual-content publishers, where substitution is genuine, may misfire if applied to discussion platforms that AI search benefits. We therefore suggest that regulatory frameworks assess harm at the level of content type and interface design rather than treating all platforms uniformly, and that they revisit these assessments as AI search shifts from static summaries toward conversational interfaces, a shift our AI Mode results show can erode complementarity.

Our study has two limitations. First, our analysis covers text-based discussion forums. Reddit is among the largest content platforms and broadly representative of online discussion, so its scale is a strength rather than a constraint. The scope limitation lies in content format: how the search-experience distinction governs the effect of AI search on other formats, such as video and music, remains an open question that we cannot address with text data. Second, our AI Mode analysis documents that conversational AI search largely attenuates the discovery effect, but the post-AI-Mode observation window is limited; longer-run data would reveal whether the attenuation stabilizes or continues toward full substitution.

Several directions for future research emerge from this work. First, extending the analysis to other content formats, such as video on YouTube and music on streaming services, would test whether the search-experience framework governs AI search effects beyond text. Second, investigating whether the increase in engagement is accompanied by changes in the quality or nature of discussions would deepen our understanding of the discovery channel. More broadly, as AI search capabilities continue to advance, understanding how each successive generation reshapes the boundary between search goods and experience goods remains an important agenda.


\printbibliography


\clearpage
\begin{APPENDICES}
\renewcommand{\thesubsection}{\thesection.\arabic{subsection}}
\renewcommand{\thetable}{\thesection.\arabic{table}}
\renewcommand{\thefigure}{\thesection.\arabic{figure}}
\renewcommand{\theequation}{\thesection.\arabic{equation}}

\section{Robustness Analysis}\label{sec:appendix_robustness}
\setcounter{table}{0}
\setcounter{figure}{0}
\setcounter{equation}{0}

\subsection{Event Study}\label{sec:appendix_event_study}

We estimate a monthly event study to examine the dynamics of the treatment effect:
\begin{equation}\label{eq:event_study}
    Y_{im} =  \sum_{t \neq -1} \beta_t \left(\text{SFW}_i \times \mathbf{1}\{m - m_0 = t\}\right) + \alpha_i + \gamma_m + \varepsilon_{im},
\end{equation}
where $i$ denotes the subreddit, $m$ denotes the calendar month, $m_0$ is the treatment month (August 2024), and $t = m - m_0$ indexes months relative to treatment. The omitted reference period is $t = -1$ (July 2024). The coefficients $\beta_t$ trace the differential change in activity between SFW and NSFW subreddits relative to the month immediately before treatment. Under the parallel trends assumption, the pre-treatment coefficients $\beta_t$ for $t < -1$ should be close to zero.

Figure~\ref{fig:event_study_main} plots the coefficients from January 2024 through December 2025. All pre-treatment coefficients are negative relative to July 2024. For comments/day, they are statistically insignificant from January through April 2024 and significantly negative in May and June; for comment authors/day, they are insignificant in January and February and significantly negative from March through June. Both series rise toward the reference after the U.S.\ launch of AI Overviews in May 2024, consistent with early treatment exposure. After treatment, the coefficients peak in February 2025, hold steady from April through July 2025, and decline after the global AI Mode launch in August 2025, becoming statistically indistinguishable from zero by November 2025 (Section~\ref{sec:aimode}).

\subsection{Pre-Trend Slope Tests}\label{sec:appendix_pretrend}

We formally test the parallel trends assumption by estimating a linear differential trend between SFW and NSFW subreddits in the pre-treatment period:
\begin{equation}\label{eq:slope_test}
    Y_{im} = \delta \left(\text{SFW}_i \times (m - m_0)\right) + \alpha_i + \gamma_m + \varepsilon_{im}, \quad \text{for } m < m_0,
\end{equation}
where $m - m_0$ is the month index relative to treatment. Under parallel trends, $\delta = 0$; a significant $\delta$ indicates a differential linear pre-trend. Table~\ref{tab:pretrend_appendix} reports the results. The estimated slope $\hat{\delta}$ is small and statistically insignificant for comments/day ($\hat{\delta} = 0.133$, $p = 0.682$) and comment authors/day ($\hat{\delta} = -0.019$, $p = 0.845$), indicating no detectable differential pre-trend.

\begin{figure}[ht!]
\centering
\caption{Monthly Event Study: Comments/Day and Comment Authors/Day}
\label{fig:event_study_main}
\vspace{0.1in}
\includegraphics[width=\textwidth]{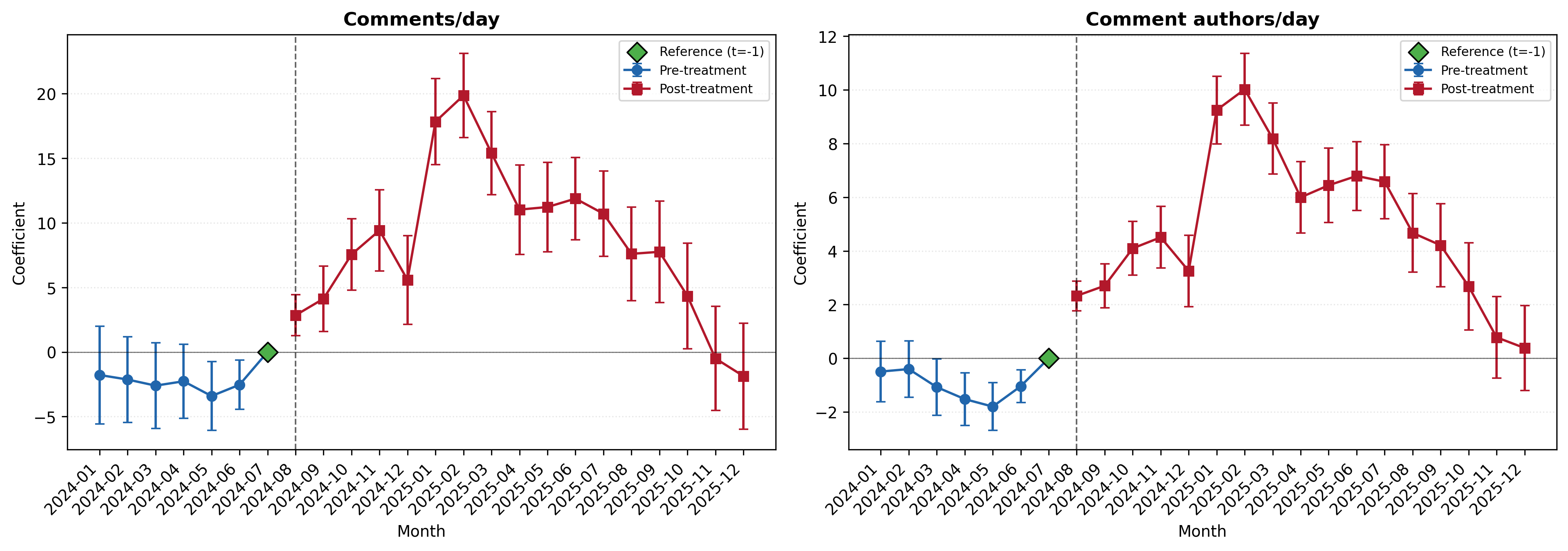}
\vspace{0.05in}
\parbox{0.85\textwidth}{\scriptsize \textit{Note}: Monthly event study coefficients from Equation~\eqref{eq:event_study}. The reference period is $t = -1$ (July 2024, green diamond). Blue circles show pre-treatment coefficients; red squares show post-treatment coefficients. Error bars indicate 95\% confidence intervals with standard errors clustered at the subreddit level.}
\end{figure}

\begin{table}[ht!]
\centering
\caption{Pre-Trend Slope Tests}
\label{tab:pretrend_appendix}
\renewcommand{\arraystretch}{1.2}
\scriptsize
\begin{threeparttable}
\begin{tabular}{l|cccl}
\hline\hline
Outcome & $\hat{\delta}$ & SE & $p$-value & Verdict \\
\hline
Comments/Day & $0.133$ & $0.326$ & $0.682$ & Pass \\
Comment Authors/Day & $-0.019$ & $0.095$ & $0.845$ & Pass \\
\hline\hline
\end{tabular}
\begin{tablenotes}
\scriptsize
\item \textit{Note}: Pre-trend slope test from Equation~\eqref{eq:slope_test}, estimated on the pre-treatment period only (January--July 2024, 7 months). $\hat{\delta}$ captures the differential linear trend for SFW relative to NSFW subreddits. Standard errors clustered at the subreddit level. ``Pass'' indicates failure to reject $H_0: \delta = 0$ at the 5\% level.
\end{tablenotes}
\end{threeparttable}
\end{table}

Table~\ref{tab:pretrend_restricted} re-estimates the pre-trend slope test from Equation~\eqref{eq:slope_test} using only the four months of January through April 2024, before Google rolled out AI Overviews in any country (during this window, AI-generated results reached only opt-in Search Labs users and a small test group of U.S.\ users).

\begin{table}[ht!]
\centering
\caption{Restricted Pre-Trend Slope Tests (January--April 2024)}
\label{tab:pretrend_restricted}
\renewcommand{\arraystretch}{1.2}
\scriptsize
\begin{threeparttable}
\begin{tabular}{l|cccl}
\hline\hline
Outcome & $\hat{\delta}$ & SE & $p$-value & Verdict \\
\hline
Comments/Day & $-0.193$ & $0.585$ & $0.742$ & Pass \\
Comment Authors/Day & $-0.375^{*}$ & $0.168$ & $0.026$ & Reject (favorable direction) \\
\hline\hline
\end{tabular}
\begin{tablenotes}
\scriptsize
\item \textit{Note}: Pre-trend slope test from Equation~\eqref{eq:slope_test} restricted to the January to April 2024 window, which precedes the general rollout of AI Overviews. $\hat{\delta}$ captures the differential linear trend for SFW relative to NSFW subreddits. Standard errors clustered at the subreddit level. ``Reject (favorable direction)'' indicates rejection of $H_0: \delta = 0$ at the 5\% level with an estimated pre-trend whose sign is opposite to that of the estimated post-treatment effect, so that any pre-trend bias works against the main finding. $^{***}p < 0.001$; $^{**}p < 0.01$; $^{*}p < 0.05$.
\end{tablenotes}
\end{threeparttable}
\end{table}

The comments slope is small and statistically insignificant, providing no evidence of a differential SFW-NSFW pre-trend in comments before the AIO rollout. The comment-authors slope is negative and significant: SFW comment-author activity was trending down relative to NSFW in the clean pre-window. Because the estimated post-treatment effect on comment authors is positive, a negative pre-trend biases against the finding, and the reported DiD estimate is a conservative lower bound on the true effect.

\subsection{HonestDiD Sensitivity Analysis}\label{sec:appendix_honestdid}

We assess the sensitivity of our results to potential violations of parallel trends using the framework of \citet{rambachan2023}. The relative magnitudes restriction bounds the maximum post-treatment trend violation as a multiple $\bar{M}$ of the largest pre-treatment shift. Figures~\ref{fig:honestdid_comments} and~\ref{fig:honestdid_comment_authors} plot the robust confidence intervals for the average treatment effect on each outcome as $\bar{M}$ increases. The breakdown value is the largest $\bar{M}$ at which the 95\% robust confidence interval still excludes zero.

For the average treatment effect, comments/day breaks down at $\bar{M} = 0.30$ (Figure~\ref{fig:honestdid_comments}) and comment authors/day at $\bar{M} = 0.50$ (Figure~\ref{fig:honestdid_comment_authors}).


\begin{figure}[ht!]
\centering
\caption{HonestDiD Sensitivity Analysis: Comments/Day (Average Effect)}
\label{fig:honestdid_comments}
\vspace{0.1in}
\includegraphics[width=0.75\textwidth]{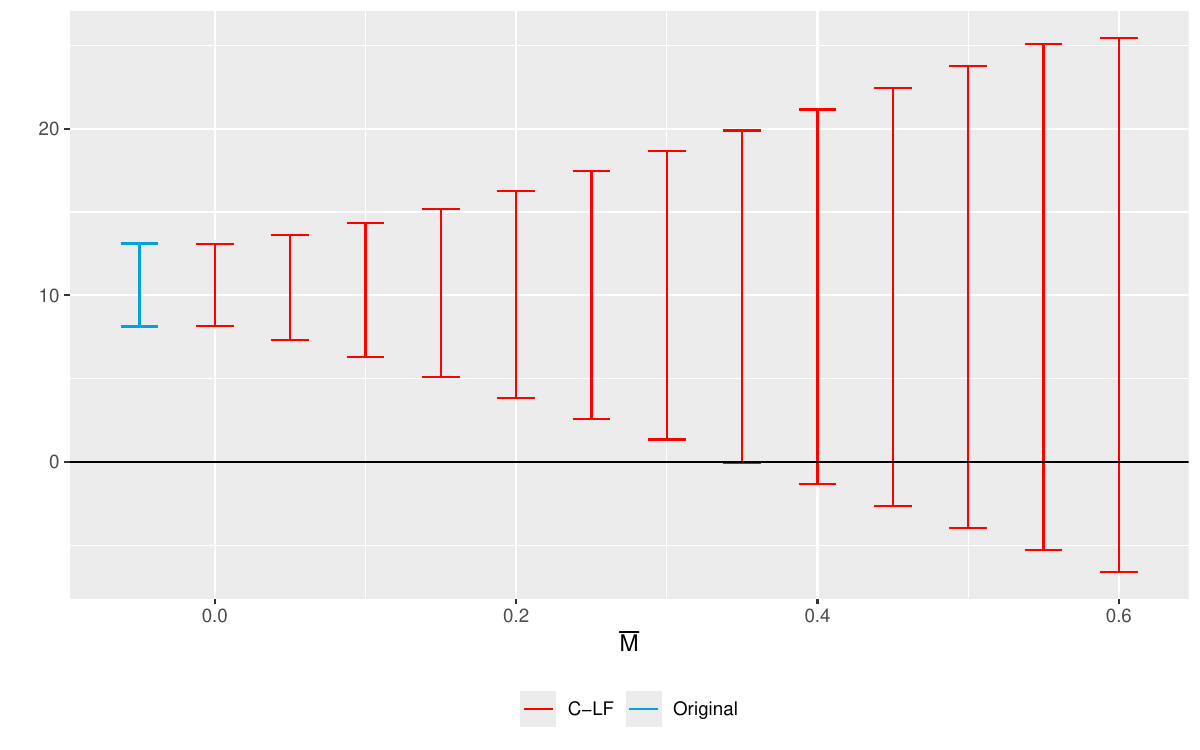}
\vspace{0.05in}
\parbox{0.85\textwidth}{\scriptsize \textit{Note}: Robust confidence intervals for the average treatment effect on comments/day under the relative magnitudes restriction \citep{rambachan2023}. The blue interval at $\bar{M} = 0$ is the original confidence interval; red intervals show the robust conditional least-favorable hybrid (C-LF) confidence intervals as $\bar{M}$ increases. The effect remains significant (CI excludes zero) for $\bar{M} \leq 0.30$.}
\end{figure}

\begin{figure}[ht!]
\centering
\caption{HonestDiD Sensitivity Analysis: Comment Authors/Day (Average Effect)}
\label{fig:honestdid_comment_authors}
\vspace{0.1in}
\includegraphics[width=0.75\textwidth]{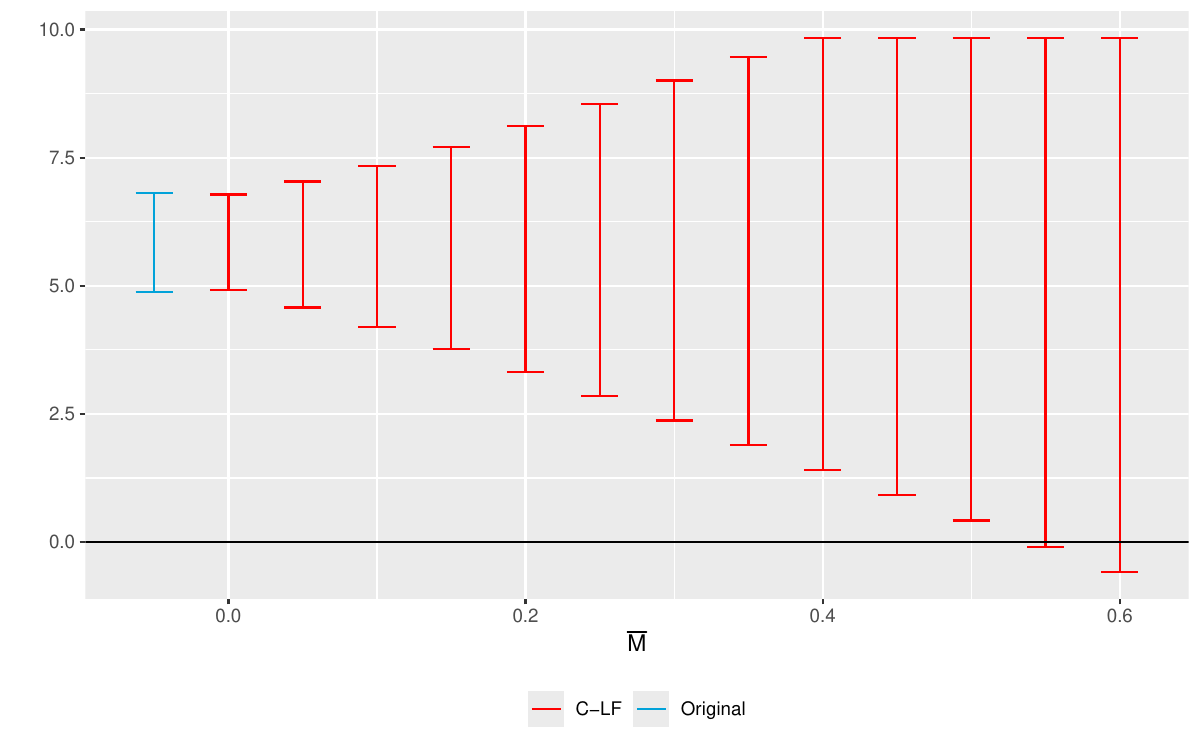}
\vspace{0.05in}
\parbox{0.85\textwidth}{\scriptsize \textit{Note}: Robust confidence intervals for the average treatment effect on comment authors/day under the relative magnitudes restriction \citep{rambachan2023}. The blue interval at $\bar{M} = 0$ is the original confidence interval; red intervals show the robust conditional least-favorable hybrid (C-LF) confidence intervals as $\bar{M}$ increases. The effect remains significant (CI excludes zero) for $\bar{M} \leq 0.50$.}
\end{figure}

\clearpage
\subsection{Alternative Treatment Dates}\label{sec:appendix_treatment_date}

AI Overviews launched to U.S. users on May 14, 2024, three months before our main treatment date of August 15, 2024 (the international expansion). Because Reddit's user base is only partly based in the United States, our pre-treatment window from January to July 2024 contains about two and a half months of partial U.S.\ exposure (May 14 to July 31, 2024), which may contaminate the pre-trend validity tests and compress the treatment-date coefficient. To assess the sensitivity of our findings to this timing issue, we (i) re-estimate the main DiD using May 14 as the treatment date and (ii) re-run the pre-trend slope test on the pre-rollout window of January to April 2024 (Appendix~\ref{sec:appendix_pretrend}). Table~\ref{tab:did_treatment_date} reports the DiD comparison.

\begin{table}[ht!]
\centering
\caption{DiD Estimates at Alternative Treatment Dates}
\label{tab:did_treatment_date}
\renewcommand{\arraystretch}{1.2}
\setlength{\tabcolsep}{6pt}
\scriptsize
\begin{threeparttable}
\begin{tabular}{ll|ccc}
\hline\hline
Treatment date & Outcome & $\hat{\beta}$ & SE & Relative effect \\
\hline
2024-05-14 (U.S.\ rollout) & Comments/Day & $10.68^{***}$ & $1.34$ & $+10.1\%$ \\
2024-05-14 (U.S.\ rollout) & Comment Authors/Day & $5.62^{***}$ & $0.50$ & $+10.5\%$ \\
\hline
2024-08-15 (international; main) & Comments/Day & $12.62^{***}$ & $1.22$ & $+12.0\%$ \\
2024-08-15 (international; main) & Comment Authors/Day & $6.67^{***}$ & $0.50$ & $+12.4\%$ \\
\hline\hline
\end{tabular}
\begin{tablenotes}
\scriptsize
\item \textit{Note}: DiD estimates from Equation~\eqref{eq:did_standard} using each treatment date on the cohort balanced panel (\num{105012} subreddits, January 2024 to July 2025). Relative effect is the coefficient divided by the pre-treatment SFW mean. Standard errors clustered at the subreddit level. $^{***}p < 0.001$; $^{**}p < 0.01$; $^{*}p < 0.05$.
\end{tablenotes}
\end{threeparttable}
\end{table}

The point estimates are positive, highly significant, and of the same sign under either treatment date. The August 15 specification yields larger coefficients for both engagement outcomes, consistent with the international rollout adding incremental effect on top of the U.S.\ launch. The qualitative conclusion that AI Overviews increase engagement on SFW subreddits relative to NSFW subreddits is robust to the choice of treatment date.




\clearpage
\subsection{Placebo Tests at Pre-AIO Policy Dates}\label{sec:appendix_placebo_policy}

Several events in the pre-AIO period could plausibly produce differential SFW versus NSFW activity trends and confound our DiD estimate. The most salient candidates are the Reddit-Google content licensing deal (February 22, 2024), which expanded SFW Reddit's visibility on Google search; the Google March 2024 Core Update (March 5, 2024), which substantially boosted Reddit's organic search traffic; and the Reddit IPO (March 21, 2024), which increased media attention to the platform. To assess whether any of these events independently produced an SFW-NSFW divergence, we re-estimate the DiD using each event date as a fake treatment date on a symmetric window of 30 days before and 30 days after the event, all of which precede the U.S.\ rollout of AI Overviews on May 14, 2024. Table~\ref{tab:placebo_policy} reports the results.

\begin{table}[ht!]
\centering
\caption{Placebo DiD Tests at Pre-AIO Policy Dates}
\label{tab:placebo_policy}
\renewcommand{\arraystretch}{1.2}
\setlength{\tabcolsep}{6pt}
\scriptsize
\begin{threeparttable}
\begin{tabular}{lll|ccc}
\hline\hline
Placebo date & Event & Outcome & $\hat{\beta}$ & SE & $p$-value \\
\hline
2024-02-22 & Reddit-Google Deal & Comments/Day & $-0.140$ & $1.023$ & $0.891$ \\
2024-03-05 & Google Core Update & Comments/Day & $+0.480$ & $0.870$ & $0.581$ \\
2024-03-21 & Reddit IPO & Comments/Day & $-0.129$ & $0.989$ & $0.896$ \\
\hline
2024-02-22 & Reddit-Google Deal & Comment Authors/Day & $-0.335$ & $0.329$ & $0.308$ \\
2024-03-05 & Google Core Update & Comment Authors/Day & $-0.368$ & $0.272$ & $0.177$ \\
2024-03-21 & Reddit IPO & Comment Authors/Day & $-0.778^{**}$ & $0.258$ & $0.003$ \\
\hline\hline
\end{tabular}
\begin{tablenotes}
\scriptsize
\item \textit{Note}: Each row reports DiD estimates from the specification in Equation~\eqref{eq:did_standard}, substituting a pre-AIO policy date for the actual August 15, 2024 treatment date. The sample window for each placebo is 30 days before and 30 days after the placebo date, all of which fall before the U.S.\ rollout of AI Overviews on May 14, 2024 and therefore precede the general rollout of AI Overviews. All specifications include subreddit and day fixed effects with standard errors clustered at the subreddit level. $^{***}p < 0.001$; $^{**}p < 0.01$; $^{*}p < 0.05$.
\end{tablenotes}
\end{threeparttable}
\end{table}

For comments per day, the placebo estimates are statistically insignificant at all three dates ($p = 0.581$ to $0.896$), indicating no differential SFW-NSFW divergence around these candidate confounders and supporting NSFW as a clean control group for comment engagement. For comment authors per day, the placebo estimates are insignificant at the Reddit-Google Deal and Google Core Update dates ($p = 0.308$ and $0.177$), and significantly negative only at the Reddit IPO date ($\hat{\beta} = -0.778$, $p = 0.003$). This residual estimate is opposite in sign to the positive AIO treatment effect of $+\num{6.667}$ for comment authors reported in Table~\ref{tab:main_did}: any pre-AIO SFW-NSFW divergence in comment authors trends downward, working against our main finding and implying that the DiD estimate for comment authors is a conservative lower bound rather than an inflated upper bound.

\clearpage
\subsection{Validity of NSFW Subreddits as a Control Group}\label{sec:appendix_nsfw_content}

A concern about NSFW as a control group is that NSFW content may be in secular decline for reasons unrelated to AI Overviews, with the recent wave of U.S. state age-verification laws targeting adult content as the leading candidate confounder. We address this concern in two ways. First, the model-free evidence in Figure~\ref{fig:model_free} shows that NSFW activity remains approximately flat throughout the post-treatment period rather than continuing to decline; if NSFW were on a sustained policy-driven downward trajectory, we would expect a monotonic decline rather than the stable series we observe. Second, the legal scope of the state age-verification laws does not reach Reddit.

By the end of our study window, more than twenty U.S. states had enacted age-verification statutes for adult content, all built around a common substantial-portion threshold: the laws apply to commercial websites where more than one-third (or a similar fraction) of the content is sexual material harmful to minors. The Supreme Court upheld the constitutionality of Texas's statute (HB 1181) in \citet{fsc_paxton_2025}. By its terms, the threshold targets predominantly adult sites: a mainstream social media platform is covered only if its own content crosses it. The dominant business response has been site-level withdrawal: Aylo, the operator of Pornhub, YouPorn, and Redtube, has blocked access to its sites in over twenty U.S. states rather than implement age verification.\footnote{See, e.g., National Today, ``Pornhub Blocked in 23 US States and 2 Countries Amid Age Verification Laws,'' February 2026, \url{https://nationaltoday.com/us/oh/wyoming-oh/news/2026/02/14/pornhub-blocked-in-23-us-states-and-2-countries-amid-age-verification-laws/}.} Table~\ref{tab:age_ver_laws} lists the major state laws active during our study window and the corresponding pattern of site blocking.

\begin{table}[ht!]
\centering
\caption{U.S. State Age-Verification Laws for Adult Content (Selected, 2023--2025)}
\label{tab:age_ver_laws}
\renewcommand{\arraystretch}{1.2}
\setlength{\tabcolsep}{6pt}
\scriptsize
\begin{threeparttable}
\begin{tabular}{lll|cc}
\hline\hline
State & Statute & Effective date & Aylo sites blocked & Reddit subject \\
\hline
Louisiana & Act 440 & Jan 2023 & No$^{a}$ & No \\
Utah & SB 287 & May 2023 & Yes & No \\
Virginia & SB 1515 & Jul 2023 & Yes & No \\
Montana & SB 544 & Jan 2024 & Yes & No \\
North Carolina & HB 8 & Jan 2024 & Yes & No \\
Texas & HB 1181 & Mar 2024 & Yes & No \\
Idaho, Kansas, Mississippi & H 498 / SB 394 / HB 1126 & Jul 2024 & Yes & No \\
Kentucky, Nebraska & HB 278 / LB 1092 & Jul 2024 & Yes & No \\
Indiana & SB 17 & Aug 2024 & Yes & No \\
Florida & HB 3 & Jan 2025 & Yes & No \\
Tennessee & SB 1792 & Jan 2025 & Yes & No \\
\hline\hline
\end{tabular}
\begin{tablenotes}
\scriptsize
\item \textit{Note}: Selected U.S. state age-verification laws active during the study window (January 2024 -- July 2025). Each statute requires commercial websites on which more than one-third (or a similar substantial portion) of content is ``sexual material harmful to minors'' to implement age verification. ``Aylo sites blocked'' indicates that Aylo (operator of Pornhub, YouPorn, and Redtube) blocked access to its sites in the listed state rather than comply, as reported in the news coverage cited in the body footnote. ``Reddit subject'' reflects the structural fact that Reddit's site-wide content mix falls below the one-third sexual-material threshold; we are aware of no enforcement action or site block targeting Reddit under any of these laws during the study window.
\item[$^{a}$] Aylo did not block Louisiana; it implemented age verification through the state's LA Wallet digital ID.
\end{tablenotes}
\end{threeparttable}
\end{table}

Reddit's content mix falls below the substantial-portion threshold by content-share measures: a publicly available estimate places NSFW posts at roughly 5\% of total Reddit submissions.\footnote{SQ Magazine, ``Reddit Statistics 2026,'' \url{https://sqmagazine.co.uk/reddit-statistics/}. Reddit does not publish a primary breakdown of NSFW share, so this figure should be treated as an approximation.} Consistent with this, Reddit has not been blocked under any U.S. state age-verification law during our study window, nor required to implement age verification in any U.S. state.\footnote{Reddit's formal age-verification deployments are confined to non-U.S.\ jurisdictions, and all but one begin after our study window: the United Kingdom (effective July 25, 2025, in the final week of the window, under the U.K.\ Online Safety Act), Australia (effective December 10, 2025, under the Social Media Minimum Age legislation), and Brazil (effective March 17, 2026, under Brazil's Digital ECA). See Reddit Help, ``Why is Reddit asking for my age?'' \url{https://support.reddithelp.com/hc/en-us/articles/36429514849428}, and Reddit Help, ``How does age verification work in Australia?'' \url{https://support.reddithelp.com/hc/en-us/articles/47163443918228}.} If anything, the displacement of users from blocked adult-content sites is more likely to inflate NSFW Reddit activity than to depress it: in states where Aylo properties are unavailable, users may substitute toward NSFW Reddit communities, which would bias our DiD estimate downward and work against the positive treatment effect we report. We therefore conclude that the wave of age-verification laws does not threaten the validity of NSFW as our control group.

\clearpage
\subsection{Alternative Cohort Definitions}\label{sec:appendix_cohort}

Table~\ref{tab:robustness_cohort_appendix} replicates the main DiD analysis using alternative cohort activity thresholds. Our main analysis uses subreddits with at least 10 submissions in December 2023. Here we also report results for thresholds of 20 and 50 minimum submissions, yielding cohorts of \num{73410} and \num{46096} subreddits, respectively, after the same filters that exclude user-profile pages, quarantined subreddits, and non-public subreddits. The results are quantitatively similar across all three thresholds, suggesting that our findings are not driven by the particular activity threshold used to define the cohort.

\begin{table}[ht!]
\centering
\caption{Robustness: Alternative Cohort Definitions}
\label{tab:robustness_cohort_appendix}
\renewcommand{\arraystretch}{1.1}
\setlength{\tabcolsep}{8pt}
\scriptsize
\begin{threeparttable}
\begin{tabular}{l|ccc|ccc}
\hline\hline
 & \multicolumn{3}{c|}{Comments/Day} & \multicolumn{3}{c}{Comment Authors/Day} \\
Cohort & $\hat{\beta}$ & SE & Effect size & $\hat{\beta}$ & SE & Effect size \\
\hline
Min 20 ($N = \num{73410}$) & \num{17.348}$^{***}$ & \num{1.745} & 11.3\% & \num{9.280}$^{***}$ & \num{0.719} & 11.9\% \\
Min 50 ($N = \num{46096}$) & \num{23.096}$^{***}$ & \num{2.610} & 9.5\% & \num{12.311}$^{***}$ & \num{1.021} & 10.0\% \\
\hline\hline
\end{tabular}
\begin{tablenotes}
\scriptsize
\item \textit{Note}: Each row reports DiD estimates from Equation~\eqref{eq:did_standard} using an alternative cohort activity threshold (minimum submissions in December 2023). The main specification (Min 10, $N = \num{105012}$) is reported in Table~\ref{tab:main_did}. All specifications include subreddit and day fixed effects with standard errors clustered at the subreddit level. Effect size is the coefficient divided by the pre-treatment SFW mean. $N$ is the number of subreddits. $^{***}p < 0.001$; $^{**}p < 0.01$; $^{*}p < 0.05$.
\end{tablenotes}
\end{threeparttable}
\end{table}

\subsection{Influential Periods and Outliers}\label{sec:appendix_influential}

Two features of the data could in principle drive the main estimates: a single unusually active calendar month and a small number of very high-activity subreddit-days. We address each with a separate specification.

\noindent \textbf{Leave-one-month-out.} As Figure~\ref{fig:model_free} shows, SFW comment activity peaks in February 2025. To verify that the main effect does not hinge on this single month, we re-estimate Equation~\eqref{eq:did_standard} after dropping all February 2025 observations from the panel. The estimates are nearly unchanged: comments/day falls from \num{12.621} to \num{11.840} and comment authors/day from \num{6.667} to \num{6.315}, with relative effects of 11.2\% and 11.8\% (versus 12.0\% and 12.4\% in the main specification).

\noindent \textbf{Winsorization.} Subreddit-day activity is highly right-skewed, so a handful of extreme observations could exert outsized leverage. We winsorize both outcomes at the 1st and 99th percentiles (pooled across all subreddit-days) and re-estimate the DiD. Because winsorizing caps the largest values, the coefficient magnitudes shrink in absolute terms (comments/day to \num{9.003}, comment authors/day to \num{4.751}), but the relative effects, measured against the correspondingly winsorized pre-treatment SFW means, remain 12.1\% and 12.0\%, essentially identical to the main estimates. The main results are therefore not an artifact of either an influential month or a small number of outlier observations.

\begin{table}[t!]
\centering
\caption{Robustness: Influential Periods and Outliers}
\label{tab:robustness_influential}
\renewcommand{\arraystretch}{1.1}
\setlength{\tabcolsep}{8pt}
\scriptsize
\begin{threeparttable}
\begin{tabular}{l|ccc|ccc}
\hline\hline
 & \multicolumn{3}{c|}{Comments/Day} & \multicolumn{3}{c}{Comment Authors/Day} \\
Specification & $\hat{\beta}$ & SE & Effect size & $\hat{\beta}$ & SE & Effect size \\
\hline
Main specification & \num{12.621}$^{***}$ & \num{1.225} & 12.0\% & \num{6.667}$^{***}$ & \num{0.496} & 12.4\% \\
Drop February 2025 & \num{11.840}$^{***}$ & \num{1.218} & 11.2\% & \num{6.315}$^{***}$ & \num{0.490} & 11.8\% \\
Winsorized (1\%/99\%) & \num{9.003}$^{***}$ & \num{0.398} & 12.1\% & \num{4.751}$^{***}$ & \num{0.186} & 12.0\% \\
\hline\hline
\end{tabular}
\begin{tablenotes}
\scriptsize
\item \textit{Note}: Each row reports DiD estimates from Equation~\eqref{eq:did_standard} under an alternative sample or outcome transformation. The main specification (Min 10, $N = \num{105012}$) is reported in Table~\ref{tab:main_did}. ``Drop February 2025'' excludes all observations in the peak month of February 2025. ``Winsorized (1\%/99\%)'' caps each outcome at its pooled 1st and 99th percentiles. All specifications include subreddit and day fixed effects with standard errors clustered at the subreddit level. Effect size is the coefficient divided by the pre-treatment SFW mean (winsorized mean for the winsorized row). $^{***}p < 0.001$; $^{**}p < 0.01$; $^{*}p < 0.05$.
\end{tablenotes}
\end{threeparttable}
\end{table}


\clearpage
\section{Subreddit Classification}\label{sec:appendix_classification}
\setcounter{table}{0}
\setcounter{figure}{0}
\setcounter{equation}{0}

\subsection{Classification Details}\label{sec:appendix_classification_details}

We classify each subreddit as a search good or an experience good using Google's Gemini 3 Flash large language model (temperature = 0 for deterministic output, thinking disabled). For each subreddit, the model receives the subreddit's name, title, and public description, drawn from the Arctic Shift Reddit Subreddits Metadata 2024-01 dump \citep{heitmann2024arcticshift}, distributed via Academic Torrents.\footnote{\url{https://academictorrents.com/details/c902f4b65f0e82a5e37db205c3405f02a028ecdf}.} The prompt frames the task in the original \citet{nelson1970} language and applies a single decision rule: could a user obtain most of the subreddit's value by reading posts alone, with no comments or interaction? It includes no reference to AI, substitution, or post-treatment outcomes, isolating the classification from the engagement metrics we study and avoiding inputs mechanically correlated with the DiD outcomes. The full prompt is reproduced verbatim below.

\begin{quote}
\small
\texttt{You are classifying Reddit subreddits using the Nelson (1970) search vs experience goods framework, adapted for online communities.}

\texttt{Core distinction:}
\begin{itemize}\setlength{\itemsep}{0pt}\setlength{\parskip}{0pt}
\small
\item \texttt{SEARCH goods: value can be judged BEFORE consumption from objective, inspectable information.}
\item \texttt{EXPERIENCE goods: value can only be judged AFTER consumption, through subjective, contextual, or interactive use.}
\end{itemize}

\texttt{Adapted to subreddits: how does a user get value from this community's content?}

\texttt{SEARCH-type subreddits: the value is in the posted content itself. A user could get most of the value by BROWSING alone. Examples: news, sports scores, factual Q\&A, product specs, image galleries, memes.}

\texttt{EXPERIENCE-type subreddits: the value is in the social interaction, interpretation, or shared context. Examples: advice-seeking, recommendations, debates, personal narratives, emotional support, hobby exchange, interactive learning.}

\texttt{Decision rule for mixed content: if a user could ONLY READ posts (no comments, no interaction), would they still get most of the subreddit's value? YES -> SEARCH; NO -> EXPERIENCE.}

\texttt{NSFW subreddits: apply the same criteria. NSFW status is irrelevant to the search/experience distinction.}

\texttt{Ambiguous subreddits: if the name and description are uninformative, assign low confidence (0.3--0.5) rather than guessing.}

\texttt{Subreddit name: r/\{subreddit\_name\}}\\
\texttt{Subreddit title: \{title\}}\\
\texttt{Public description: \{public\_description\}}

\texttt{Respond with ONLY a JSON object:}\\
\texttt{\{``classification'': ``search'' or ``experience'', ``confidence'': 0.0--1.0, ``reason'': ``one sentence citing what in the name/description drove the decision''\}}

\texttt{Confidence scale: 0.9--1.0 clearly one type; 0.7--0.9 likely with explicit signals; 0.5--0.7 mixed; 0.3--0.5 weak signal; 0.0--0.3 no usable signal.}
\end{quote}

The model returns a binary classification, a confidence score on a 0.0--1.0 scale, and a brief justification. We process all \num{105012} subreddits in the cohort using 20 parallel API requests with deterministic decoding. Of these, \num{104975} return a parseable label; 37 cases either fail to parse or are blocked by the model's safety filter. Table~\ref{tab:classification_summary} summarizes the resulting classification.

\begin{table}[ht!]
\centering
\caption{Subreddit Classification Summary}
\label{tab:classification_summary}
\renewcommand{\arraystretch}{1.1}
\setlength{\tabcolsep}{8pt}
\scriptsize
\begin{threeparttable}
\begin{tabular}{l|rrr}
\hline\hline
& Search & Experience & Total \\
\hline
SFW (treatment) & \num{34134} & \num{36540} & \num{70674} \\
NSFW (control) & \num{25316} & \num{8985} & \num{34301} \\
\hline
Total & \num{59450} & \num{45525} & \num{104975} \\
\hline\hline
\end{tabular}
\begin{tablenotes}
\scriptsize
\item \textit{Note}: Classification counts by treatment group (SFW/NSFW) and content type (search/experience). NSFW subreddits are disproportionately classified as search goods (73.8\%), consistent with NSFW content being primarily image/video consumption rather than discussion. SFW subreddits split more evenly (51.7\% experience). Of the \num{104975} classified subreddits, \num{85332} (81.3\%) receive high-confidence scores ($\geq 0.8$).
\end{tablenotes}
\end{threeparttable}
\end{table}

\subsection{High-Confidence Classification Robustness}\label{sec:appendix_highconf}

To address the concern that the triple-difference estimate may be driven by ambiguously classified subreddits, we re-estimate Equation~\eqref{eq:triple_diff} on the subsample of \num{85332} subreddits (81.3\% of the cohort) that receive a Gemini confidence score of at least 0.8. Table~\ref{tab:triple_diff_highconf} reports the results. The triple-difference coefficient $\hat{\beta}_3$ remains highly significant for both outcomes and is comparable in magnitude to the full-sample estimate in Table~\ref{tab:triple_diff}, indicating that the differential effect of AI Overviews on experience-good communities is not driven by borderline classifications.

\begin{table}[ht!]
\centering
\caption{Triple-Difference: High-Confidence Classifications Only ($\geq 0.8$)}
\label{tab:triple_diff_highconf}
\renewcommand{\arraystretch}{1.1}
\setlength{\tabcolsep}{8pt}
\scriptsize
\begin{threeparttable}
\begin{tabular}{l|cc}
\hline\hline
& Comments/Day & Comment Authors/Day \\
\hline
SFW $\times$ Post & $9.45^{***}$ & $4.49^{***}$ \\
& $(1.813)$ & $(0.786)$ \\[4pt]
Exp $\times$ Post & $-0.69$ & $-0.78^{\dagger}$ \\
& $(1.961)$ & $(0.450)$ \\[4pt]
SFW $\times$ Post $\times$ Exp & $9.57^{**}$ & $6.15^{***}$ \\
& $(3.049)$ & $(1.203)$ \\[4pt]
\hline
Implied experience DiD ($\beta_1 + \beta_3$) & $19.02$ & $10.64$ \\
Experience / Search ratio & $2.0\times$ & $2.4\times$ \\
\hline
Subreddit FE & Yes & Yes \\
Day FE & Yes & Yes \\
Subreddits & \num{85332} & \num{85332} \\
Observations & \num{49321896} & \num{49321896} \\
\hline\hline
\end{tabular}
\begin{tablenotes}
\scriptsize
\item \textit{Note}: Same specification as Equation~\eqref{eq:triple_diff}, restricted to subreddits with Gemini confidence $\geq 0.8$. Standard errors clustered at the subreddit level in parentheses. $^{***}p < 0.001$; $^{**}p < 0.01$; $^{*}p < 0.05$; $^{\dagger}p < 0.1$.
\end{tablenotes}
\end{threeparttable}
\end{table}

\clearpage
\section{Validation of Google AI Overview Content Policy}\label{sec:appendix_aio_validation}
\setcounter{table}{0}
\setcounter{figure}{0}
\setcounter{equation}{0}

Our identification strategy relies on the assumption that Google AI Overviews reference SFW Reddit content but not NSFW Reddit content. We validate this assumption through a manual audit of Google search results.

\subsection{Sampling Procedure}

We construct a sample of 1,000 Reddit post titles by collecting 5 titles each from 100 SFW subreddits and 100 NSFW subreddits.

\paragraph{SFW sample.} We use Reddit's \texttt{r/popular} feed, which serves as the default front page for logged-out users and excludes all NSFW communities, to identify the 100 SFW subreddits with the most trending posts. For each of the 100 SFW subreddits, we collect the 5 most trending posts, excluding moderator-pinned posts such as community announcements and rules, yielding 500 SFW post titles.

\paragraph{NSFW sample.} Because \texttt{r/popular} excludes NSFW communities, we obtain the NSFW subreddit list from a publicly available directory that ranks NSFW subreddits by subscriber count (\url{https://postpone.app/list-of-nsfw-subreddits}) and select the top 100 NSFW communities from this list. For each subreddit, we query the ``hot'' feed (\url{https://www.reddit.com/r/{subreddit}/hot.json}), skip community announcements and rule posts, and take the top 5 trending posts, yielding 500 NSFW post titles.

\subsection{Search Procedure}    

We manually type each of the 1,000 post titles into the Google Chrome search bar and record (i) whether the query triggers an AI Overview and (ii) whether the AI Overview references any SFW or NSFW Reddit content. We conduct searches by manual entry rather than through a third-party API or programmatic URL construction because Google distinguishes between the two: a query typed into Chrome generates a URL with additional parameters that signal an organic user session, whereas a programmatic request lacks these parameters. Google may serve different results, including different AI Overview behavior, depending on these signals. Manual entry ensures that our audit replicates the search experience of actual users.

\subsection{Results}

Table~\ref{tab:aio_validation} summarizes the results. Among the 500 SFW post titles, 285 (57.0\%) trigger an AI Overview, and all 285 reference SFW content. Among the 500 NSFW post titles, 190 (38.0\%) trigger an AI Overview,\footnote{The 190 NSFW post titles that trigger an AI Overview do so because Google interprets the query as a non-explicit search intent: most NSFW subreddit posts pair a descriptive title with media content, so the title alone often reads as a generic query and triggers AI Overviews.} but none of the 190 summaries reference any NSFW content. All AI Overviews triggered by NSFW post titles reference only SFW sources. This confirms that Google's content moderation policy effectively prevents NSFW Reddit content from appearing in AI Overview summaries, supporting the validity of our identification strategy.

\begin{table}[ht!]
\centering
\caption{Validation of Google AI Overview Content Policy}
\label{tab:aio_validation}
\renewcommand{\arraystretch}{1.2}
\scriptsize
\begin{threeparttable}
\begin{tabular}{l|cccc}
\hline\hline
Sample & Posts searched & AIO triggered & AIO with NSFW references & AIO with SFW references \\
\hline
SFW post titles & 500 & 285 (57.0\%) & 0 & 285 \\
NSFW post titles & 500 & 190 (38.0\%) & 0 & 190 \\
\hline\hline
\end{tabular}
\begin{tablenotes}
\scriptsize
\item \textit{Note}: Manual audit conducted on April 2, 2026. SFW sample: 5 most trending posts from each of the top 100 SFW subreddits on \texttt{r/popular}. NSFW sample: 5 most trending posts from each of the top 100 NSFW subreddits. Each post title was entered manually into the Google search bar. ``AIO triggered'' indicates the query produced an AI Overview. No AI Overview referenced NSFW Reddit content in either sample.
\end{tablenotes}
\end{threeparttable}
\end{table}

\end{APPENDICES}

\end{document}